\documentclass[12pt]{article}
\usepackage{amssymb,amsmath,epsfig}

\begin{document}

\title{\textbf{Black Hole Evaporation in a Noncommutative
Charged Vaidya Model}}

\author{M. Sharif \thanks{msharif.math@pu.edu.pk} and Wajiha Javed \\
\\Department of Mathematics, University of the Punjab,\\Quaid-e-Azam
Campus Lahore-54590, Pakistan.}
\date{}

\maketitle
\begin{abstract}
The aim of this paper is to study the black hole evaporation and
Hawking radiation for a noncommutative charged Vaidya black hole.
For this purpose, we determine spherically symmetric charged Vaidya
model and then formulate a noncommutative
Reissner-Nordstr$\ddot{o}$m-like solution of this model which leads
to an exact $(t-r)$ dependent metric. The behavior of temporal
component of this metric and the corresponding Hawking temperature
is investigated. The results are shown in the form of graphs.
Further, we examine the tunneling process of the charged massive
particles through the quantum horizon. It is found that the
tunneling amplitude is modified due to noncommutativity. Also, it
turns out that black hole evaporates completely in the limits of
large time and horizon radius. The effect of charge is to reduce the
temperature from maximum value to zero. It is mentioned here that
the final stage of black hole evaporation turns out to be a naked
singularity.
\end{abstract}

\section{Introduction}

Classically, the concept of smooth spacetime manifold breaks down at
short distances. Noncommutative (NC) geometry gives an impressive
structure to investigate short distance spacetime dynamics. In this
structure, there exists a universal minimal length scale,
$\sqrt{\sigma}$ (equivalent to Planck length). In General Relativity
(GR), the effects of NC can be taken into account by keeping the
standard form of the Einstein tensor and using the altered form of
the energy-momentum tensor in the field equations. This involves
distribution of point-like structures in favor of smeared objects
\footnote{An object constructed by means of generalized function
$\rho(t,r)$ is smeared in space and is known as smeared object.
These objects are non-local. However, smearing cannot change the
physical nature of the object but the spatial structure of the
object is changed, which is smeared in a certain region determined
by $\sqrt{\sigma}$.}. Noncommutative black holes (BHs) require an
appropriate framework in which the NC corresponds to GR.

Black hole evaporation leads to comprehensive and straightforward
predictions for the distribution of emitted particles. However, its
final phase is unsatisfactory and cannot be resolved due to
semi-classical representation of Hawking process. Black Hole
evaporation can be explored in curved spacetime by quantum field
theory but BH itself is described by a classical background
geometry. On the other hand, the final stage of BH decay requires
quantum gravity corrections while the semi-classical model is
incapable to discuss evaporation. Noncommutative quantum field
theory (based on the coordinate coherent states) treats
short-distance behavior of point-like structures, where mass and
charge are distributed throughout a region of size $\sqrt{\sigma}$.

Hawking \cite{7} suggested that a radiation spectrum of an
evaporating BH is just like a black body with a purely thermal
spectrum, i.e., BH can radiate thermally. Consequently, a
misconception \cite{N1} was developed with respect to information
loss from BH leading to non-unitary of quantum evolution
{\footnote{Non-unitary quantum evolution is one of the
interpretations of information paradox to modify quantum mechanics.
In unitary evolution, entropy is constant with usual $S$-matrix
whereas it is not constant in non-unitary quantum evolution.}.
Accordingly, when a BH evaporates completely, all the information
related to matter, falling inside the BH, will be lost. Gibbons and
Hawking \cite{N2} proposed a formulation to visualize radiation as
tunneling of charged particles. In this formulation, radiation
corresponds to electron-positron pair creation in a constant
electric field whereas the energy of a particle changes sign as it
crosses the horizon. The total energy of a pair, created just inside
or outside the horizon, would be zero when one member of the pair
tunneled to opposite side. Parikh and Wilczek \cite{8} derived
Hawking radiation as a tunneling through the quantum horizon on the
basis of null geodesics. In this framework, the corrected BH
radiation spectrum is obtained due to back-reaction effects. This
tunneling process shows that the extended radiation spectrum is not
exactly thermal yielding a unitary quantum evolution.

There are two different semiclassical tunneling methods to calculate
the tunneling amplitude which leads to the Hawking temperature. The
first method, called the null geodesic method, gives the same
temperature as the Hawking temperature. The second one named as
canonically invariant tunneling, leads to canonically invariant
tunneling amplitude and hence the corresponding temperature which is
higher than the Hawking temperature by a factor of 2 \cite{NN2}.
Akhmedova \emph{et al} \cite{NN1} argued that a particular
coordinate transformation resolves this problem in quasiclassical
picture.

Alexeyev \emph{et al} \cite{NN3} discussed BH evaporation spectra in
Einstein-dilaton-Gauss-Bonnet four dimensional string gravity model
by using the radial null geodesic method. They showed that BHs
should not disappear and become relics at the end of the evaporation
process. They investigated numerically the possibility of
experimental detection of such remnant BHs and discussed mass loss
rate in analytic form. These primordial BH relics could form a part
of the non-baryonic dark matter in our universe.

Smailagic and Spallucci \cite{2} found various NC models in terms of
coordinate coherent states which satisfy Lorentz invariance,
unitarity and UV finiteness of quantum field theory. Nicolini et al.
\cite{3} derived the generalized NC metric which does not allow BH
to decay lower than a minimal nonzero mass $M_0$, i.e., BH remnant
mass. The effects of NC BHs have been studied \cite{N4,N5} and found
consistent results. The evaporation process stops when a BH
approaches to a Planck size remnant with zero temperature. Also, it
does not diverge rather reaches to a maximum value before shrinking
to an absolute zero temperature which is an intrinsic property of
manifold. Some other people \cite{20} also explored information loss
problem during BH evaporation.

Sharif and Javed \cite{M1} investigated quantum corrections of the
thermodynamical quantities for a Bardeen charged regular BH by using
quantum tunneling approach over semiclassical approximations. In a
recent work \cite{M2}, they have also discussed the behavior of NC
on the thermodynamics of this BH. Mehdipour \cite{35.2} analyzed the
tunneling of massive particles through quantum horizon of the NC
Schwarzschild BH and derived the modified Hawking radiation,
thermodynamical quantities and emission rate. He also discussed
stable BH remnant and information loss issues. Nozari and Mehdipour
\cite{35.1} studied the effects of smeared mass and showed that
information might be saved by a stable BH remnant during the
evaporation process. Mehdipour \cite{P3} extended this work for NC
Reissner-Nordstr$\ddot{o}$m (RN) BH and determined the emission rate
consistent with unitary theory. The same author \cite{P2} also
formulated a NC Schwarzschild-like metric for a Vaidya solution and
analyzed three possible causal structures of BH initial and remnant
mass. Also, he studied the tunneling of charged particles across the
quantum horizon of the Schwarzschild-like Vaidya BH and evaluated
the corresponding entropy.

The purpose of this paper is two fold: Firstly, we formulate NC
RN-like solution of the spherically symmetric charged Vaidya model.
Secondly, we investigate some of its features. In particular, we
explore BH evaporation and Parikh-Wilczek tunneling process. Its
format is as follows. In section \textbf{2}, we solve the coupled
field equations for spherically symmetric charged Vaidya model. The
effect of NC form of this model is investigated in the framework of
coordinate coherent states in section \textbf{3}. Here, an exact
$(t-r)$ dependent RN-like BH solution is obtained. Section
\textbf{4} yields the behavior of the temporal component of this
solution and also provides discussion about BH evaporation in the
limits of large time and charge. In section \textbf{5}, we study
Parikh-Wilczek tunneling for such a Vaidya solution and also Hawking
temperature in the presence of charge. The tunneling amplitude at
which massless particles tunnel across the event horizon is
computed. Finally, the conclusion of the work is given in the last
section. It is mentioned here that throughout the paper we assume
$\hbar=c=G=1$.

\section{Charged Vaidya Model}

This section is devoted to formulate spherically symmetric charged
Vaidya model in the RN-like form by using the procedure given by
Farley and D'Eath \cite{P1}. Here we shall skip the details of the
procedure as it is already available and use only the required
results. The spherically symmetric Vaidya form metric is given by
(2.34) of \cite{P1}
\begin{equation}
ds^2=-e^{\nu(t,r)}dt^2+e^{\mu(t,r)}dr^2+r^2d\Omega^2,\label{2.34}
\end{equation}
where
\begin{equation*}
e^{\nu(t,r)}=\left(\frac{\dot{M}}{\chi(M)}\right)^2e^{-\mu},\quad
e^{-\mu(t,r)}=1-\frac{2M}{r}, \quad
d\Omega^2=d\theta^2+\sin^2\theta d\phi^2,
\end{equation*}
$M(t,r)$ is a slowly-varying mass function and $\chi(M)$ depends
on the details of the radiation. The corresponding field equations
are \cite{P1}
\begin{eqnarray}
&&\mu^\prime=8\pi rT_{rr}+\frac{1-e^\mu}{r},\label{3-1}\\
&&\nu^\prime=8\pi re^{\mu-\nu}T_{tt}-\frac{1-e^\mu}{r},\label{3-2}\\
&&\dot{\mu}=8\pi rT_{tr},\label{3-3}\\
&&1-e^{-\mu}+\frac{1}{2}re^{-\mu}(\mu^\prime-\nu^\prime)-\frac{1}{2}r^2R^{(0)}=8\pi
T_{\theta\theta}=\frac{8\pi
T_{\phi\phi}}{\sin^2\theta}\label{3-4},
\end{eqnarray}
where
\begin{eqnarray}
R^{(0)}=-8\pi
{T^a}_a=-\frac{2}{r^2}(1-e^{-\mu})+e^{-\frac{1}{2}(\mu+\nu)}
[{(\dot{\mu}e^{\frac{1}{2}(\mu-\nu)})}^{\cdot}-{(\nu^\prime
e^{\frac{1}{2}(\nu-\mu)})}^{\prime}]\label{3-5}.
\end{eqnarray}
Here dot and prime mean derivatives with respect to time and $r$
respectively. It is mentioned here that Eqs.(\ref{3-1}) and
(\ref{3-3}) represent the Hamiltonian and momentum constraints
respectively \cite{1-10}. Equations (\ref{3-1}) and (\ref{3-2}) lead
to
\begin{equation}
\frac{1}{2}(\mu^\prime-\nu^\prime)=\frac{1-e^\mu}{r}\label{3-7}
\end{equation}
while Eqs.(\ref{3-4}) and (\ref{3-5}) yield
\begin{equation}
T_{rr}=e^{(\mu-\nu)}T_{tt}.\label{3-6}
\end{equation}

For the spherically symmetric Vaidya metric of the form (\ref{2.34}), we
define $e^{-\mu(t,r)}$ by adding charge $Q(t,r)$ as follows
\begin{equation}
e^{-\mu(t,r)}=1-\frac{2M(t,r)}{r}+\frac{Q^2(t,r)}{r^2}.\label{3-8}
\end{equation}
Using the procedure \cite{P1}, one can write from the field
equations
\begin{equation}
{T^t}_r e^{\frac{1}{2}(\nu-\mu)}+ {T^t}_t
=0.\label{3-13}
\end{equation}
Also, using Eqs.(\ref{3-1}), (\ref{3-3}), (\ref{3-6}) and
(\ref{3-13}), we obtain
\begin{equation}
\mu^\prime+\frac{e^\mu-1}{r}+\dot{\mu}e^{\frac{(\mu-\nu)}{2}}=0.\label{3-14}
\end{equation}
Inserting the value of $e^\mu$ from Eq.(\ref{3-8}), it follows that
\begin{equation}
e^{\nu(t,r)}=\left(\frac{\frac{2Q\dot{Q}}{r}-2\dot{M}}{2M^\prime
-\frac{2QQ^\prime}{r}+\frac{Q^2}{r^2}}\right)^2\left(1-\frac{2M}{r}
+\frac{Q^2}{r^2}\right)^{-1}.\label{3-16}
\end{equation}
The corresponding form of the Vaidya solution \cite{1-7} will become
\begin{eqnarray}
ds^2&=&-\left(\frac{\frac{2Q\dot{Q}}{r}-2\dot{M}}{2M^\prime
-\frac{2QQ^\prime}{r}+\frac{Q^2}{r^2}}\right)^2\left(1-\frac{2M}{r}
+\frac{Q^2}{r^2}\right)^{-1}dt^2\nonumber\\
&+&\left(1-\frac{2M}{r}+\frac{Q^2}{r^2}\right)^{-1}dr^2
+r^2d\Omega^2.\label{3-18}
\end{eqnarray}
This is the spherically symmetric charged Vaidya model.

Now we transform this metric so that it is in RN-like form. For this
purpose, we write Eq.(\ref{3-16}) in the following form
\begin{equation}
e^{\frac{(\nu-\mu)}{2}}=\frac{r(2Q\dot{Q}-2r\dot{M})}{2r^2M^\prime-2rQQ^\prime+Q^2}.
\label{3-15}
\end{equation}
Differentiating Eq.(\ref{3-15}) with respect to $r$ and using
Eq.(\ref{3-7}), it follows that
\begin{eqnarray}
&&\frac{2Mr-Q^2}{r^2-2Mr+Q^2}{[(Q\dot{Q}-r\dot{M})(2r^2M^\prime
-2rQQ^\prime+Q^2)]}\nonumber\\
&&=(2r^2M^\prime-2rQQ^\prime
+Q^2)(rQ\dot{Q}^\prime+rQ^\prime\dot{Q}+Q\dot{Q}
-r^2\dot{M}^\prime\nonumber\\
&&-2\dot{M}r)-(2rQ\dot{Q}-2r^2\dot{M})
(r^2M^{\prime\prime}+2M^\prime r-rQQ^{\prime\prime}-r{Q^\prime}^2)\nonumber
\end{eqnarray}
which can also be written as
\begin{equation}
\frac{{[(2M^\prime-\frac{2QQ^\prime}{r}+\frac{Q^2}{r^2})(1-\frac{2M}{r}
+\frac{Q^2}{r^2})]}^\cdot}{{[(2M^\prime-\frac{2QQ^\prime}{r}+\frac{Q^2}{r^2})
(1-\frac{2M}{r}+\frac{Q^2}{r^2})]}^\prime}=\frac{(2M
-\frac{Q^2}{r}\dot{)}}{(2M-\frac{Q^2}{r})'}.\label{3-21}
\end{equation}
This has the solution
\begin{equation}
\left(2M^\prime-\frac{2QQ^\prime}{r}+\frac{Q^2}{r^2}\right)
\left(1-\frac{2M}{r}+\frac{Q^2}{r^2}\right)=\chi(M,Q),\label{3-22}
\end{equation}
where $\chi(M,Q)\geq0$. With the help of this equation, we can write
Eq.(\ref{3-16}) as follows
\begin{equation}
e^{\nu(t,r)}=e^{2\Psi(t,r)}\left(1-\frac{2M(t,r)}{r}+\frac{Q^2(t,r)}{r^2}\right),
\label{3-23}
\end{equation}
where
\begin{equation*}
e^{2\Psi(t,r)}=\left(\frac{-\frac{2Q\dot{Q}}{r}+2\dot{M}}{\chi(M,Q)}\right)^2.
\end{equation*}
Consequently, the line element (\ref{2.34}) turns out to be
\begin{eqnarray}
ds^2&=&-e^{2\Psi(t,r)}\left(1-\frac{2M(t,r)}{r}+\frac{Q^2(t,r)}{r^2}\right)dt^2
\nonumber\\
&+&\left(1-\frac{2M(t,r)}{r}+\frac{Q^2(t,r)}{r^2}\right)^{-1}dr^2
+r^2d\Omega^2.\label{3-25}
\end{eqnarray}
This is the required spherically symmetric charged Vaidya model in
RN-like form. For a specific choice of
$\chi(M,Q)=-(-\frac{2Q\dot{Q}}{r}+2\dot{M}),~\Psi(t,r)$ vanishes and
hence it reduces to simple RN-like form, i.e.,
\begin{equation}
ds^2=-F(t,r)dt^2+F^{-1}(t,r)dr^2+r^2d\Omega^2,\label{2}
\end{equation}
where $F=1-\frac{2M(t,r)}{r}+\frac{Q^2(t,r)}{r^2}$.

\section{Noncommutative Black Hole}

Here we develop NC form of the RN-like Vaidya metric (\ref{2}) by
using the coordinate coherent states formalism \cite{P2}. The
mass/energy and charge distribution can be written by the following
smeared delta function $\rho$ \cite{P3,P2}
\begin{eqnarray}
\rho_{matt}(t,r)&=&\frac{M}{(4\pi\sigma)^{\frac{3}{2}}}e^{-\frac{r^2}{4\sigma}},\label{1}\\
\rho_{el}(t,r)&=&\frac{Q}{(4\pi\sigma)^{\frac{3}{2}}}e^{-\frac{r^2}{4\sigma}},\label{1.5}
\end{eqnarray}
respectively, where $\sigma$ is the NC factor. The energy-momentum
tensor for self-gravitating and anisotropic fluid source is given by
\begin{eqnarray}
{T_a}^b=\left(\begin{array}{cccc}
{T_t}^t &{T_t}^r &0&0\\
{T_r}^t &{T_r}^r &0&0\\
0&0&{T_\theta}^\theta&0\\
0&0&0&{T_\phi}^\phi\\
\end{array}\right).\label{9}\\
\nonumber
\end{eqnarray}
Here we take ${T_t}^t=-(\rho_{matt}+\rho_{el})={T_r}^r$. The
corresponding field equations become
\begin{eqnarray}
F^\prime r+F+8\pi r^2 (\rho_{matt}+\rho_{el})-1&=&0,\label{5}\\
\dot{F}-8\pi rF^2 {T_t}^r&=&0,\label{6}\\
rF''F^3-2r\dot{F}^2+r\ddot{F}F+2F^3F'-16\pi
rF^3{T_\theta}^{\theta}&=&0,\label{7}\\
{T_t}^r={T_r}^t,\quad{T_\phi}^\phi&=&{T_\theta}^\theta.\label{8}
\end{eqnarray}

The conservation of energy-momentum tensor, ${{T_a}^b}_{;b}=0$,
yields
\begin{eqnarray*}
\partial_t{T_t}^t+\partial_r{T_r}^r+\frac{1}{2}g^{tt}\partial_rg_{tt}({T_r}^r-{T_t}^t)
-\frac{1}{2}g^{rr}\partial_tg_{rr}({T_r}^r
-{T_t}^t)&&\\+g^{\theta\theta}\partial_r
g_{\theta\theta}({T_r}^r-{T_\theta}^\theta)&=&0
\end{eqnarray*}
which leads to
\begin{eqnarray*}
\partial_t{T_t}^t+\partial_r{T_t}^t
+g^{\theta\theta}\partial_rg_{\theta\theta}({T_t}^t-{T_\theta}^\theta)&=&0.\label{3}
\end{eqnarray*}
Inserting the values, we obtain
\begin{equation}
{T_\theta}^\theta=(\rho_{matt}+\rho_{el})\left(\frac{-r(\dot{M}+\dot{Q}
+M^\prime+Q^\prime)}{2(M+Q)}+\frac{r^2}{4\sigma}-1\right).\label{4}
\end{equation}
Now we consider the perfect fluid condition at large distances to
determine the mass and charge functions. For this purpose, we take
isotropic pressure terms, i.e., ${T_r}^r={T_\theta}^\theta$ so
that the above equation yields
\begin{equation}
M+Q=Ce^{[\frac{t^2}{4\sigma}+\frac{t(r-t)}{2\sigma}]},\label{10}
\end{equation}
where $C(r-t)$ is an integration function and can be defined as
\begin{equation}
C(r-t)=M_I-\frac{Q_I^2\left[\varepsilon^2\left(\frac{r-t}{2\sqrt{\sigma}}\right)
-\frac{1}{\sqrt{\pi}}\left(\frac{r-t}{\sqrt{2\sigma}}\right)\varepsilon
\left(\frac{r-t}{\sqrt{2\sigma}}\right)\right]}{2r\left[\varepsilon
\left(\frac{r-t}{2\sqrt{\sigma}}\right)\left(1+\frac{t^2}{2\sigma}\right)
-\frac{r}{\sqrt{\pi\sigma}}e^{-\frac{(r-t)^2}{4\sigma}}\left(1+\frac{t}{r}
\right)\right]}.\label{10.5}
\end{equation}
Here $M_I$ and $Q_I$ are initial BH mass and charge respectively
while the Gauss error function is defined as
$\varepsilon(x)\equiv\frac{2}{\sqrt{\pi}}\int_0^xe^{-p^2}dp$. Using
Eq.(\ref{10}) in (\ref{5}), it follows that
\begin{equation}
F(t,r)=1-\frac{2M_\sigma(t,r)}{r}+\frac{Q_\sigma^2(t,r)}{r^2},\label{12}
\end{equation}
where the Gaussian smeared mass and charge distribution are
\begin{eqnarray}
M_\sigma(t,r)&=&M_I\left[\varepsilon\left(\frac{r-t}{2\sqrt{\sigma}}\right)
\left(1+\frac{t^2}{2\sigma}\right)-\frac{re^{-
\frac{(r-t)^2}{4\sigma}}}{\sqrt{\pi\sigma}}
\left(1+\frac{t}{r}\right)\right],\nonumber\\
Q_\sigma(t,r)&=&Q_I\left[\varepsilon^2\left(\frac{r-t}{2\sqrt{\sigma}}\right)-
\frac{1}{\sqrt{\pi}}\left(\frac{r-t}{\sqrt{2\sigma}}\right)\varepsilon
\left(\frac{r-t}{\sqrt{2\sigma}}\right)\right]^{\frac{1}{2}}.\label{13.5}
\end{eqnarray}
This is the NC form of (\ref{2}).

The asymptotic form of (\ref{12}) reduces to the RN metric for large
distances at $t=0$. The metric (\ref{12}) characterizes the geometry
of a NC inspired RN-like Vaidya BH. The radiating behavior of such a
modified BH can now be investigated by plotting $g_{tt}$ for
different values of $M_I$ and $Q_I$. Coordinate NC leads to the
existence of different causal structures, i.e., non-extremal BH
(with two horizon), extremal BH (with one horizon) and charged
massive droplet (with no horizon). Thus the NC BH can shrink to the
minimal nonzero mass with minimal nonzero horizon radius.

\section{Horizon Radius and Black Hole Evaporation}

Here we investigate some features of the NC metric (\ref{12}).
Firstly, we analyse the temporal component $g_{tt}=F(t,r)$ in the
form of graphs versus horizon radius $\frac{r}{\sqrt{\sigma}}$. The
following table \cite{P2} provides values of the minimal nonzero
mass as well as horizon radius with increasing time for an extremal
BH. This shows that the minimal nonzero mass decreases whereas the
minimal nonzero horizon radius increases with time indicating micro
BH evaporates completely, i.e., $M_0\rightarrow 0$ as
$\frac{t}{\sqrt{\sigma}}\gg 1$. Consequently the concept of BH
remnant does not exist.
\begin{center}
{\bf {\small Table 1.}} {\small Extremal black hole}

\vspace{0.5cm}

\begin{tabular}{|l|l|l|l|l|l|l|l|l|}
\hline {\bf Time} & {\bf Minimal nonzero}&{\bf Minimal nonzero}\\&
\textbf{mass} & \textbf{horizon radius}
\\ \hline {\bf $t=0$} &$M_0\approx1.90\sqrt{\sigma}$ & $r_0\approx3.02\sqrt{\sigma}$
\\ \hline {\bf $t=1.00\sqrt{\sigma}$}&$M_0\approx1.68\sqrt{\sigma}$&
$r_0\approx4.49\sqrt{\sigma}$
\\ \hline {\bf $t=2.00\sqrt{\sigma}$}&$M_0\approx0.99\sqrt{\sigma}$&
$r_0\approx5.34\sqrt{\sigma}$
\\ \hline {\bf $t=3.00\sqrt{\sigma}$}&$M_0\approx0.62\sqrt{\sigma}$&
$r_0\approx6.14\sqrt{\sigma}$
\\ \hline {\bf $t=4.00\sqrt{\sigma}$}&$M_0\approx0.43\sqrt{\sigma}$&
$r_0\approx7.18\sqrt{\sigma}$
\\ \hline {\bf $t=5.00\sqrt{\sigma}$}&$M_0\approx0.32\sqrt{\sigma}$&
$r_0\approx8.32\sqrt{\sigma}$
\\ \hline {\bf $t=10.00\sqrt{\sigma}$}&$M_0\approx0.13\sqrt{\sigma}$&
$r_0\approx13.27\sqrt{\sigma}$
\\ \hline {\bf $t=100.00\sqrt{\sigma}$}&$M_0\approx0.01\sqrt{\sigma}$&
$r_0\approx105.05\sqrt{\sigma}$
\\ \hline {\bf $t\longrightarrow\infty$}&$M_0\longrightarrow0$&$r_0\longrightarrow\infty$
\\ \hline
\end{tabular}
\end{center}
The graphs of $F$ are drawn for the following three cases:
$$(i)\quad M_I>M_0,\quad (ii)\quad M_I=M_0, \quad (iii)\quad M_I<M_0.$$

For the first case, the graphs of $F$ are shown in Figures
\textbf{1-3}. Here we choose different values of time
$\frac{t}{\sqrt{\sigma}}$ with $M_I>M_0$ and fixed
$\frac{M_I}{\sqrt{\sigma}}$ (i.e., $M_I=3.00\sqrt{\sigma}$). The
curves are marked from top to bottom of the right side for $t=0$,
$1.00\sqrt{\sigma}$, $2.00\sqrt{\sigma}$, $3.00\sqrt{\sigma}$ and
$4.00\sqrt{\sigma}$ respectively. This demonstrates that distance
between the horizons increases with time. When
$t\longrightarrow\infty$, we have two different horizons for the
three possibilities of initial mass and initial charge, i.e.,
$\frac{Q_I}{\sqrt{\sigma}} <\frac{M_I}{\sqrt{\sigma}}$,
$\frac{Q_I}{\sqrt{\sigma}}=\frac{M_I}{\sqrt{\sigma}}$ and
$\frac{Q_I}{\sqrt{\sigma}}>\frac{M_I}{\sqrt{\sigma}}$.
\begin{figure}\center
\epsfig{file=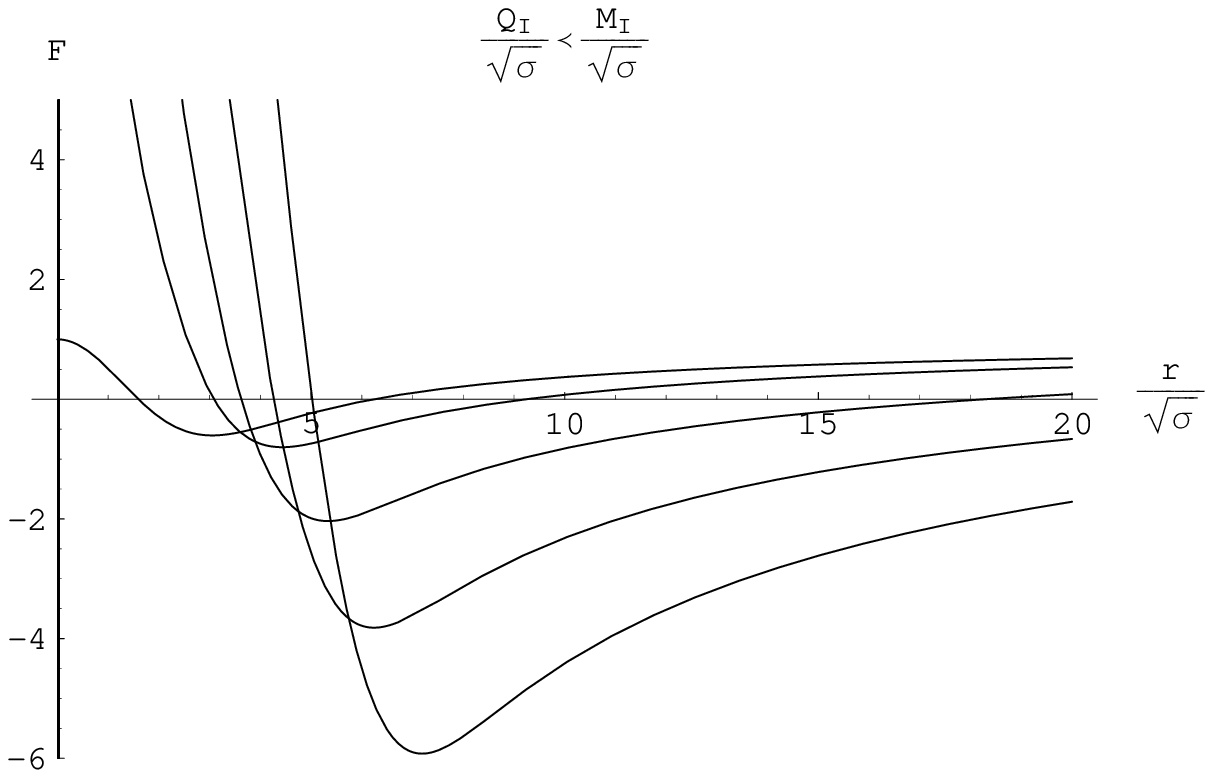, width=0.50\linewidth}\\
\caption{Here $Q_I=1.00\sqrt{\sigma} < M_I=3.00\sqrt{\sigma}$.}
\epsfig{file=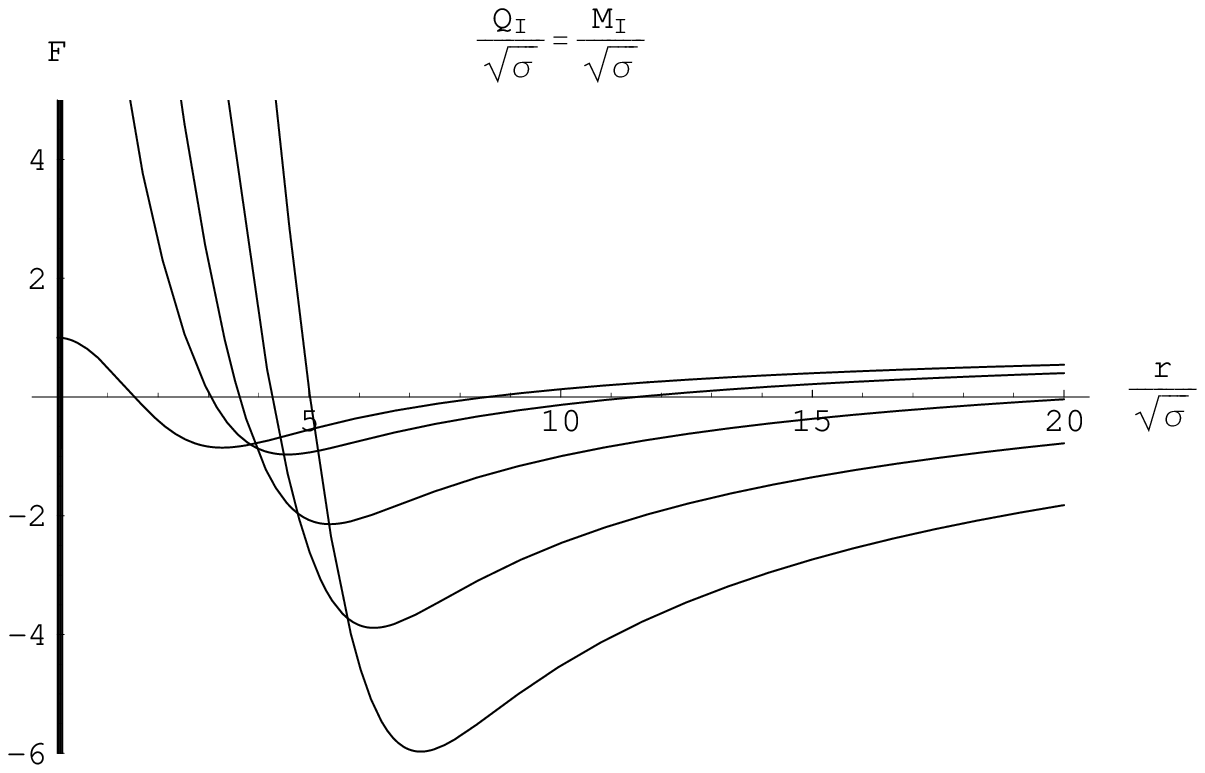, width=0.50\linewidth}\\
\caption{Here $Q_I=3.00\sqrt{\sigma} = M_I$.} \epsfig{file=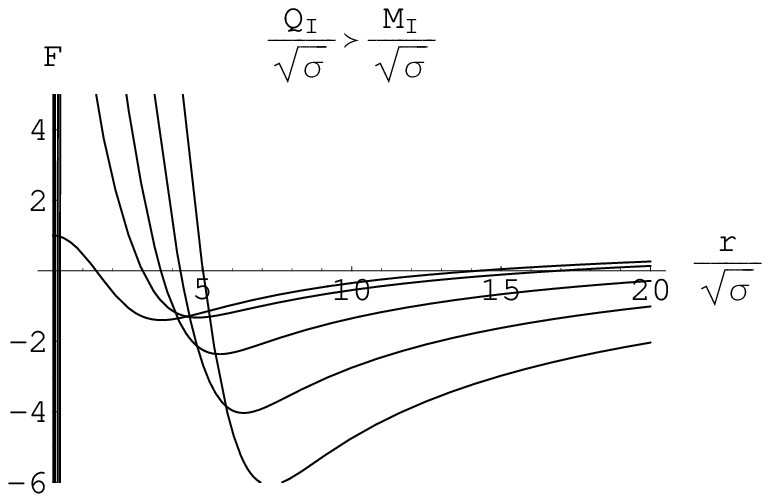,
width=0.50\linewidth} \caption{Here $Q_I=5.00\sqrt{\sigma} >
M_I=3.00\sqrt{\sigma}$.}
\end{figure}
\begin{figure}\center
\epsfig{file=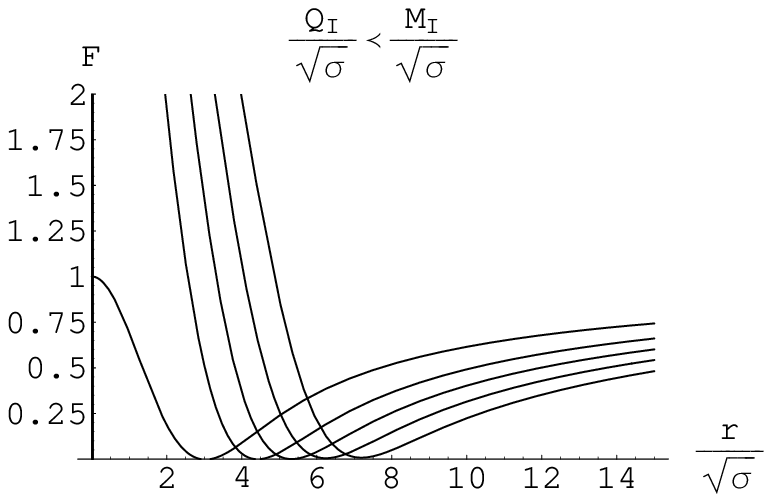, width=0.50\linewidth}\\
\caption{Here $Q_I=0.40\sqrt{\sigma} < M_I=M_0$.}
\epsfig{file=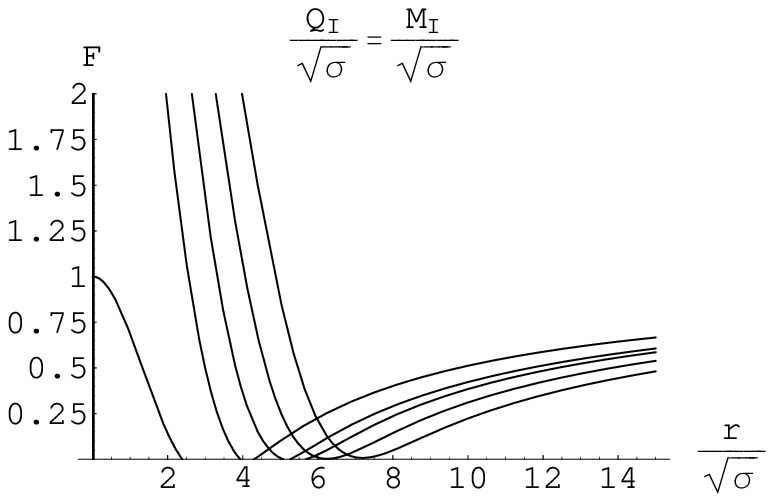, width=0.50\linewidth}\\
\caption{Here $Q_I = M_I=M_0$.} \epsfig{file=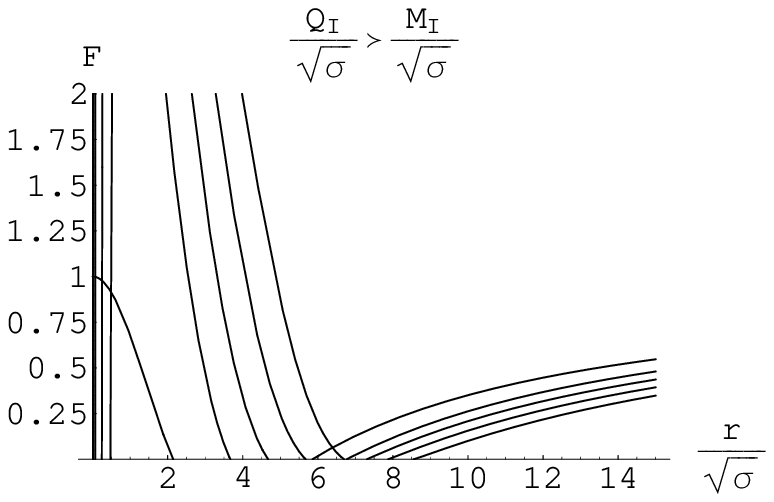,
width=0.50\linewidth} \caption{Here $Q_I=3.00\sqrt{\sigma} >
M_I=M_0$.}
\end{figure}
\begin{figure}\center
\epsfig{file=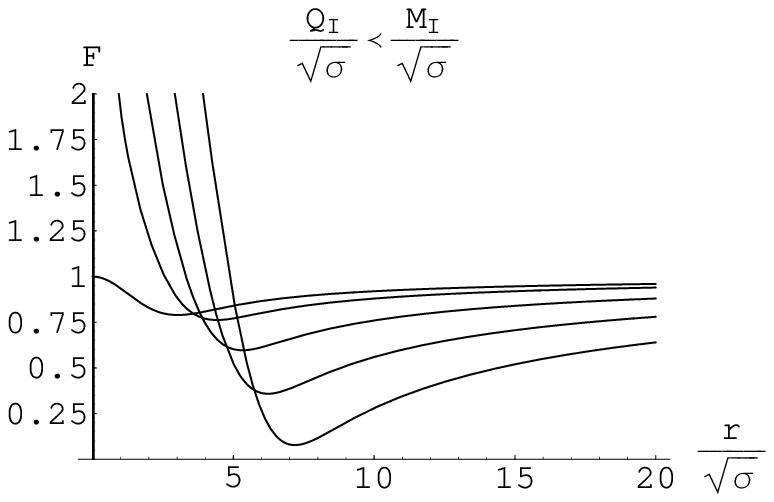, width=0.50\linewidth}\\
\caption{Here $Q_I=0.10\sqrt{\sigma} < M_I$.}
\epsfig{file=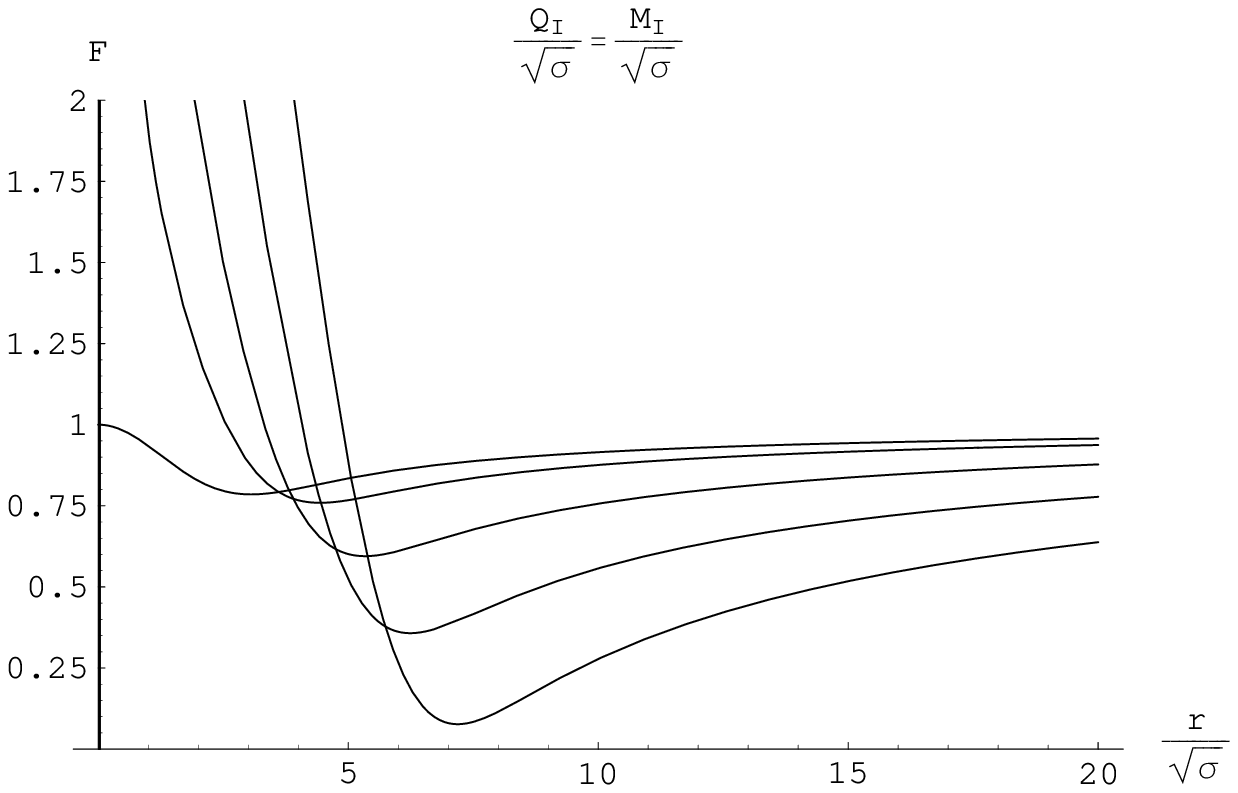, width=0.50\linewidth}\\
\caption{Here $Q_I = M_I$.} \epsfig{file=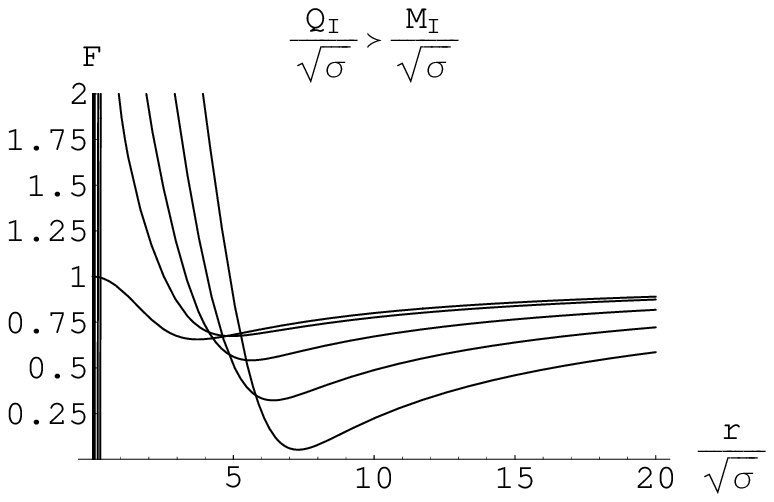,
width=0.50\linewidth} \caption{Here $Q_I=2.00\sqrt{\sigma} > M_I$.}
\end{figure}
Figures \textbf{4-6} show the graphs of $F$ when $M_I=M_0$. These
represent the possibility of an extremum structure with one
degenerate event horizon in the presence of charge. The
possibilities of $\frac{M_I}{\sqrt{\sigma}}$ and
$\frac{Q_I}{\sqrt{\sigma}}$ are given as follows:
\begin{itemize}
\item For $\frac{Q_I}{\sqrt{\sigma}}<\frac{M_I}{\sqrt{\sigma}}$ ,
it is possible to have one degenerate event horizon (extremal BH)
for all $t$.
\item For $\frac{Q_I}{\sqrt{\sigma}}=\frac{M_I}{\sqrt{\sigma}}$, there
is one degenerate event horizon for $t>2$.
\item $\frac{Q_I}{\sqrt{\sigma}}>\frac{M_I}{\sqrt{\sigma}}$, it is
impossible to have a degenerate event horizon for all $t$.
\end{itemize}
Figures \textbf{7-9} yield graphs of $F$ for the case $M_I<M_0$ with
$M_I=0.40\sqrt{\sigma}$. For all the three possibilities of initial
mass and initial charge, curves do not show any event horizon with
the passage of time.

For $t=0$, $F$ in Eq.(\ref{12}) takes the form
\begin{eqnarray}
F(r)&=&1-\frac{2M_I}{r}\left[\varepsilon\left(\frac{r}{2\sqrt{\sigma}}\right)
-\frac{r}{\sqrt{\pi\sigma}}
e^{-\frac{r^2}{4\sigma}}\right]\nonumber\\
&+&\frac{{Q_I}^2}{r^2}\left[\varepsilon^2\left(\frac{r}{2\sqrt{\sigma}}\right)-
\frac{1}{\sqrt{\pi}}\left(\frac{r}{\sqrt{2\sigma}}\right)\varepsilon
\left(\frac{r}{\sqrt{2\sigma}}\right)\right].\label{15}
\end{eqnarray}
In the commutative limit, i.e., $\sigma\longrightarrow0$, we have
$\varepsilon(x)\longrightarrow1$ and hence $F$ reduces to
\begin{equation}
F(r)=1-\frac{2M_I}{r}+\frac{{Q_I}^2}{r^2}.\label{16}
\end{equation}

\begin{figure}
\begin{tabular}{c}
\epsfig{file=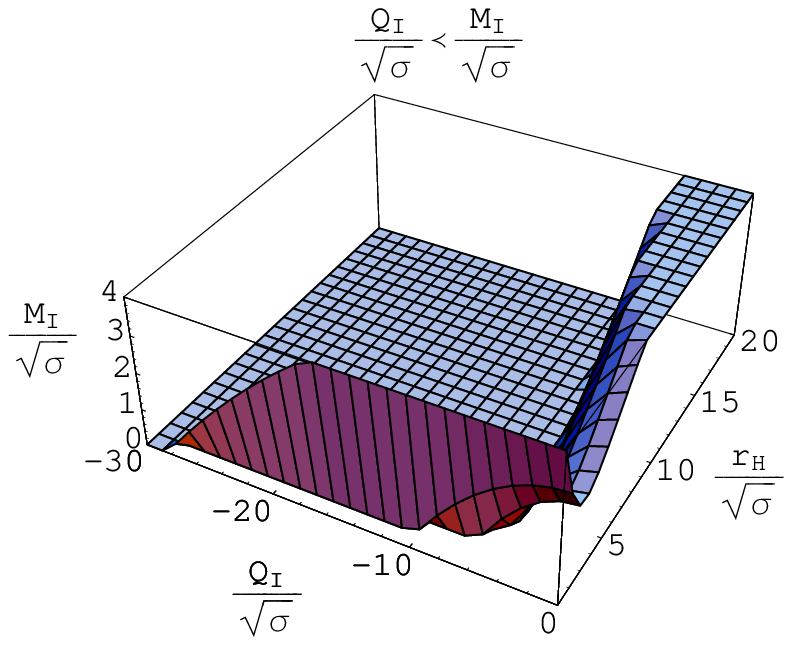, width=0.34\linewidth}
\epsfig{file=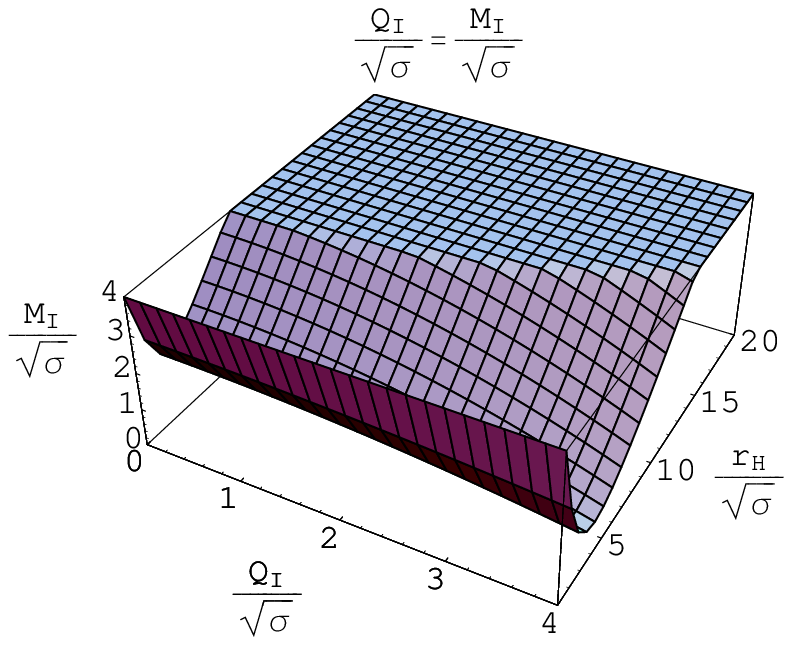, width=0.34\linewidth}
\epsfig{file=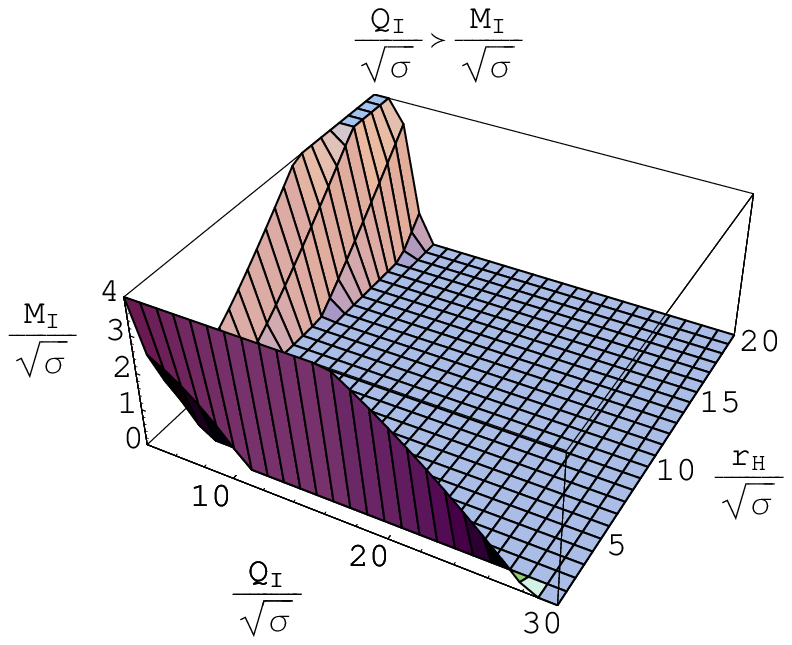, width=0.34\linewidth}\\
\end{tabular}
\caption{$\frac{M_I}{\sqrt{\sigma}}$ versus
$\frac{r_H}{\sqrt{\sigma}}$ for $t=0$.}
\begin{tabular}{c}
\epsfig{file=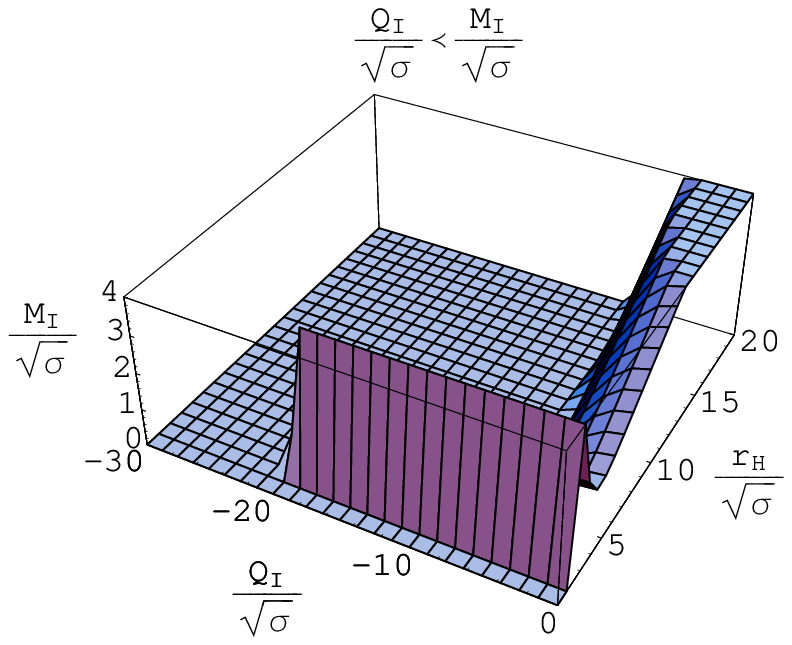, width=0.34\linewidth}
\epsfig{file=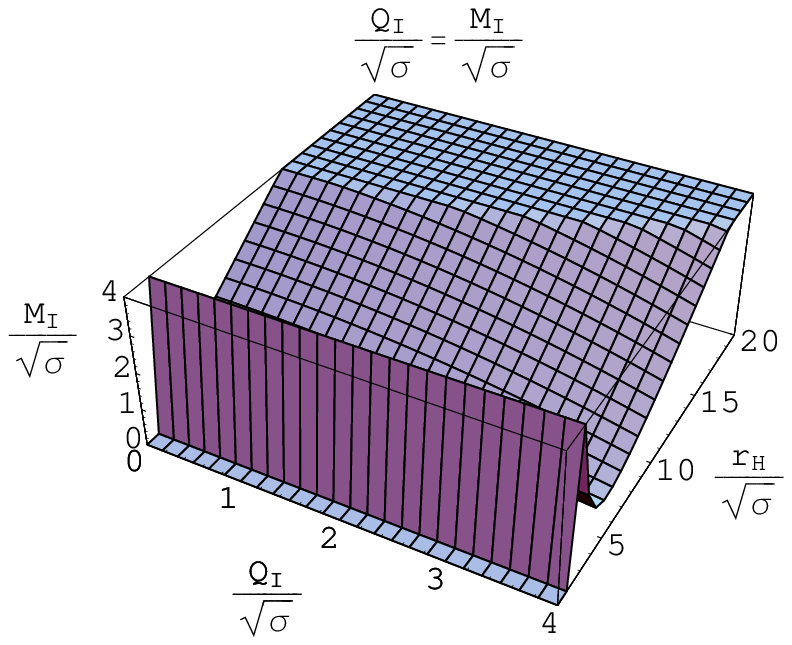, width=0.34\linewidth}
\epsfig{file=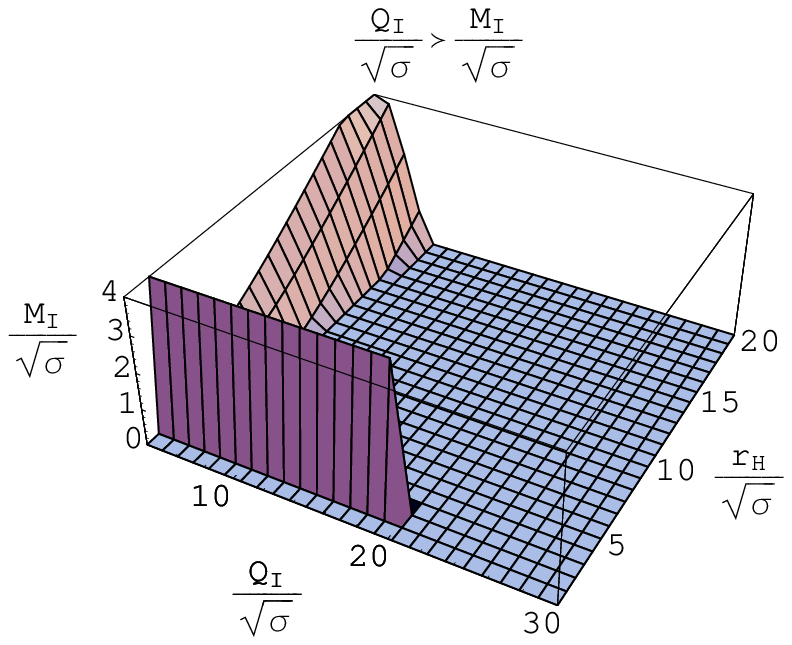, width=0.34\linewidth}\\
\end{tabular}
\caption{$\frac{M_I}{\sqrt{\sigma}}$ versus
$\frac{r_H}{\sqrt{\sigma}}$ for $t=1.00\sqrt{\sigma}$.}
\end{figure}

\begin{figure}
\begin{tabular}{c}
\epsfig{file=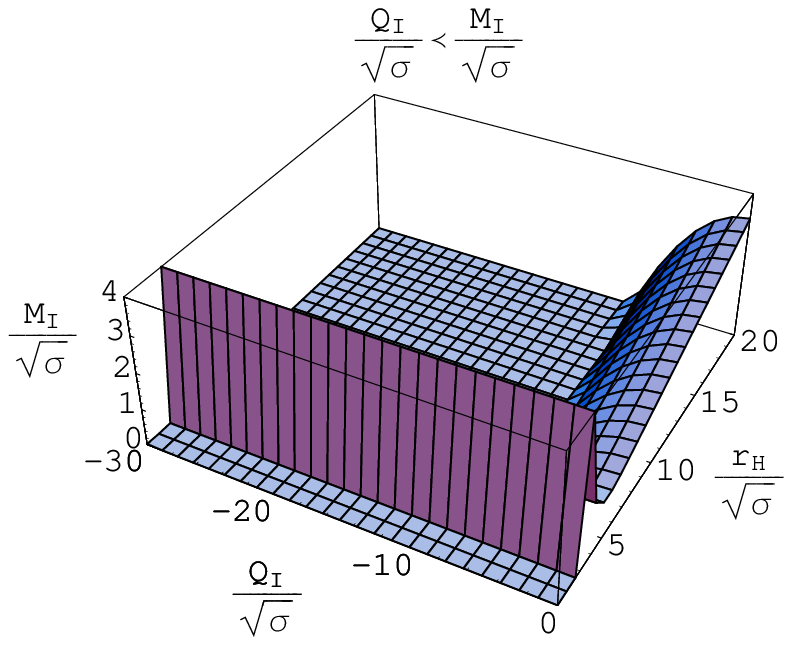, width=0.34\linewidth}
\epsfig{file=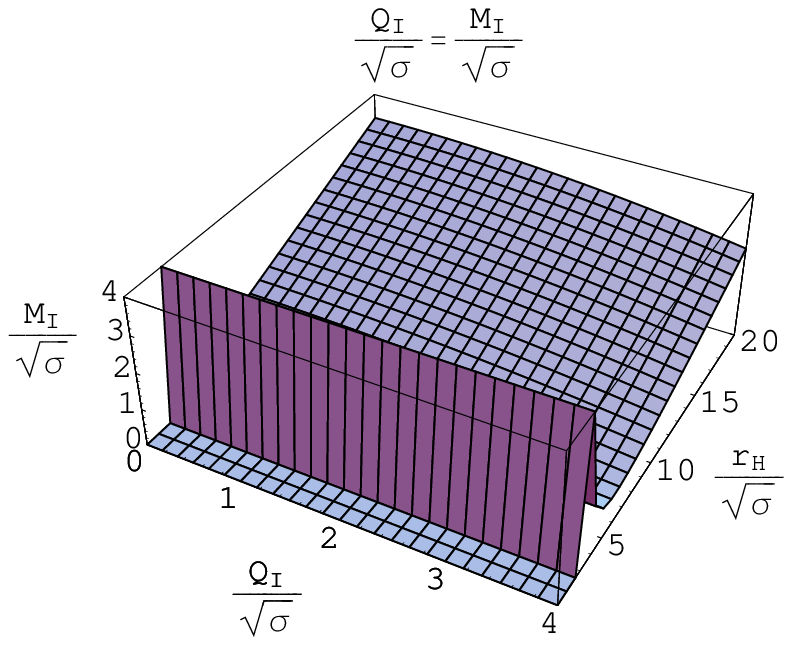, width=0.34\linewidth}
\epsfig{file=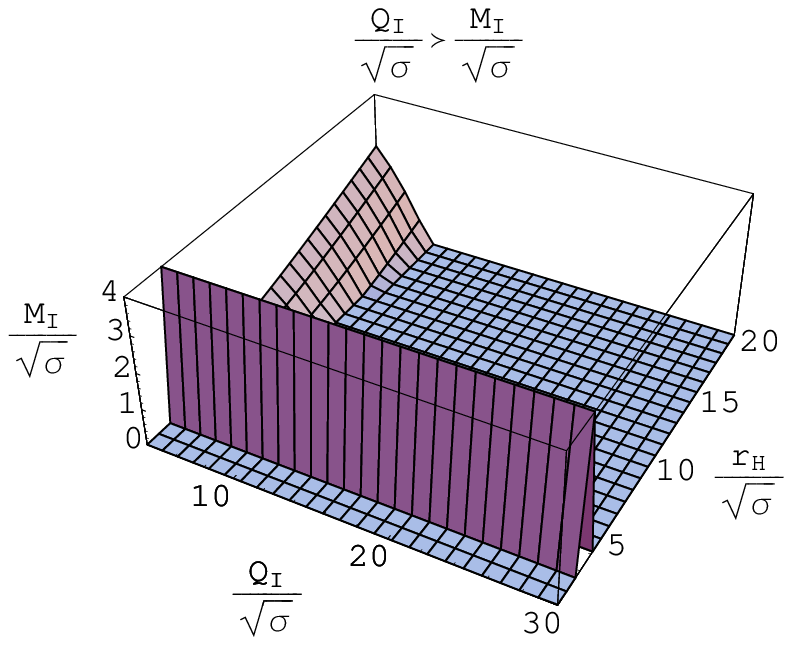, width=0.34\linewidth}
\end{tabular}
\caption{$\frac{M_I}{\sqrt{\sigma}}$ versus
$\frac{r_H}{\sqrt{\sigma}}$ for $t=2.00\sqrt{\sigma}$.}
\begin{tabular}{c}
\epsfig{file=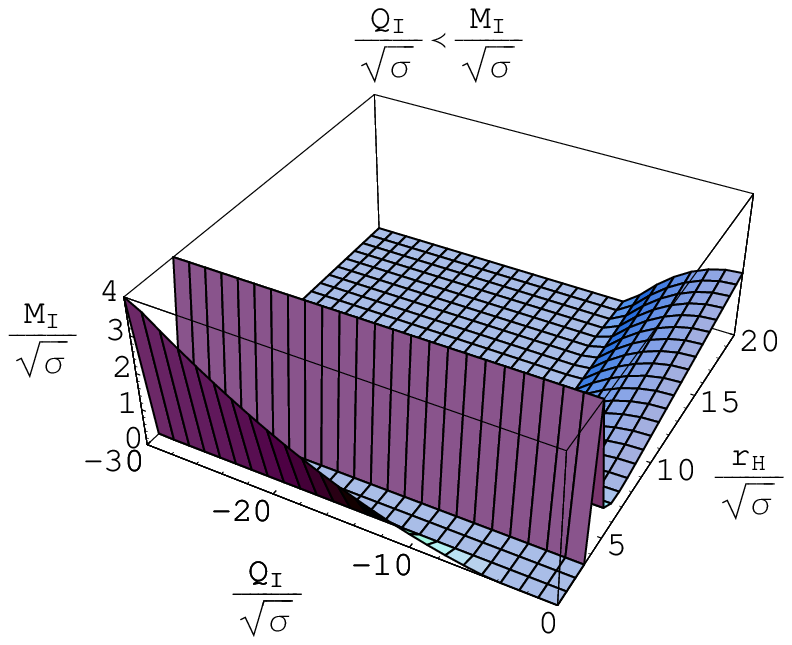, width=0.34\linewidth}
\epsfig{file=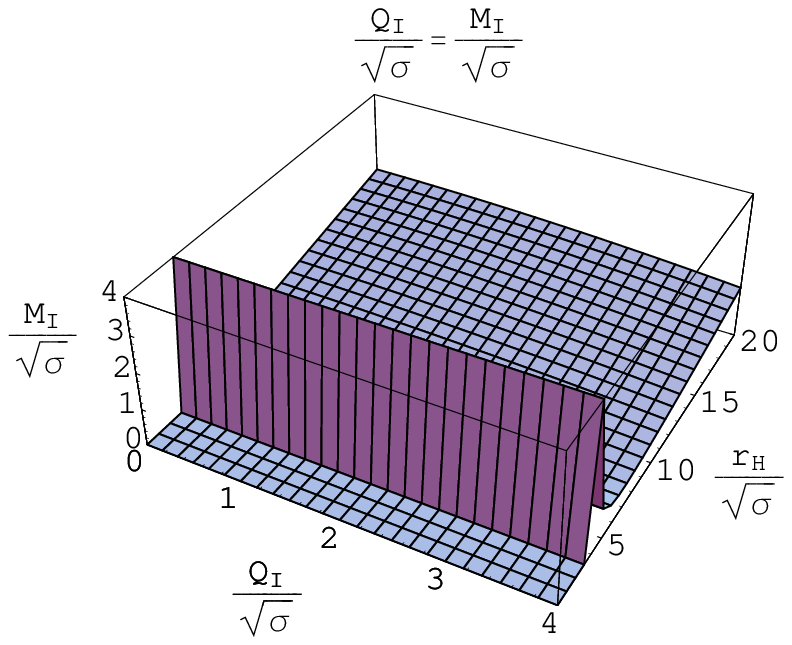, width=0.34\linewidth}
\epsfig{file=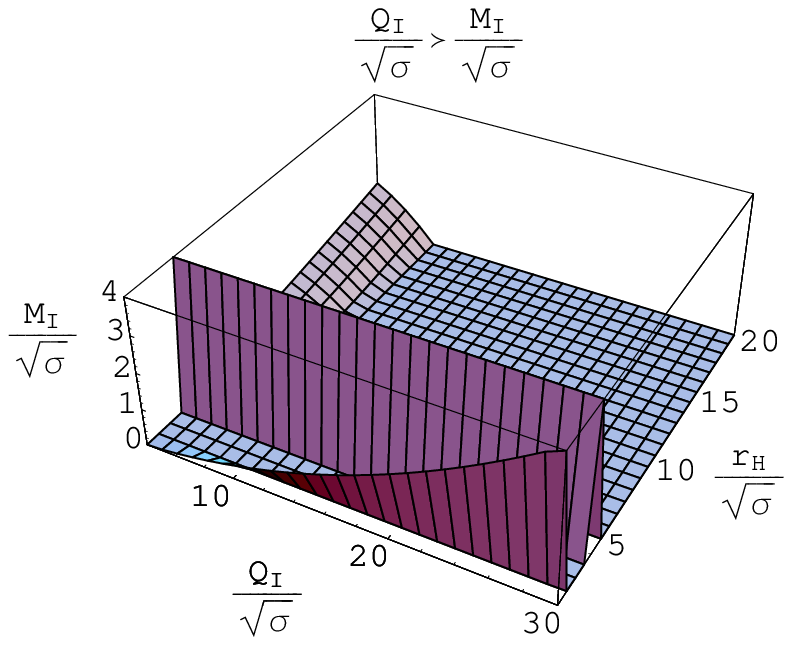, width=0.34\linewidth}\\
\end{tabular}
\caption{$\frac{M_I}{\sqrt{\sigma}}$ versus
$\frac{r_H}{\sqrt{\sigma}}$ for $t=3.00\sqrt{\sigma}$.}
\begin{tabular}{c}
\epsfig{file=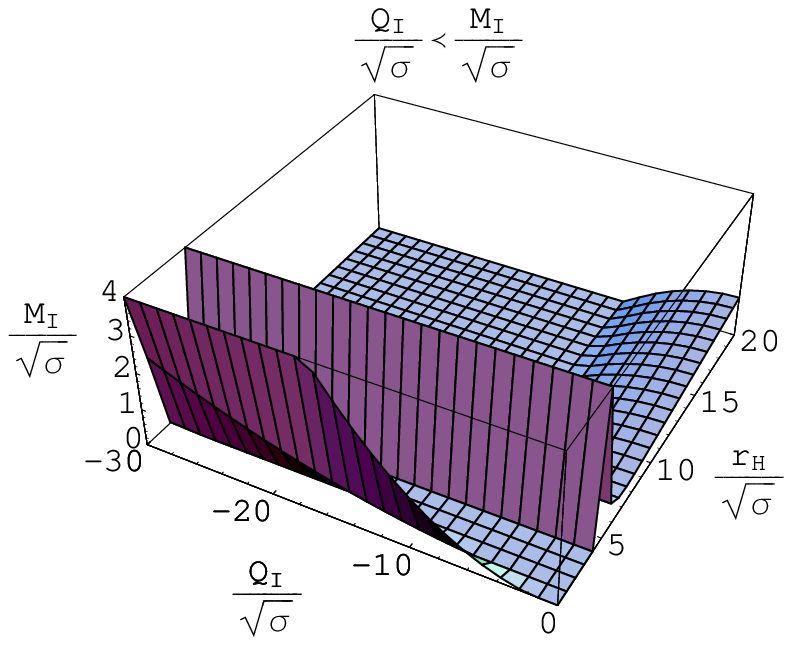, width=0.34\linewidth}
\epsfig{file=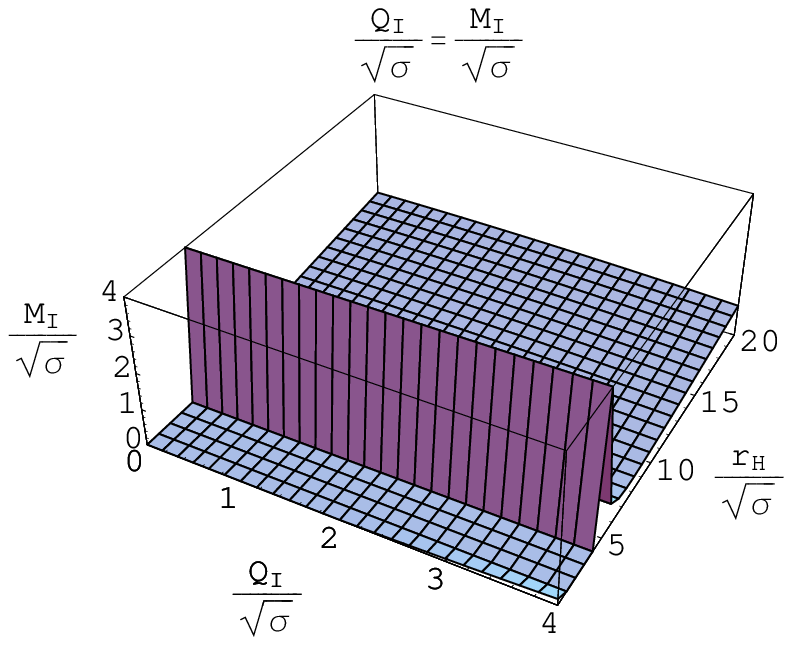, width=0.34\linewidth}
\epsfig{file=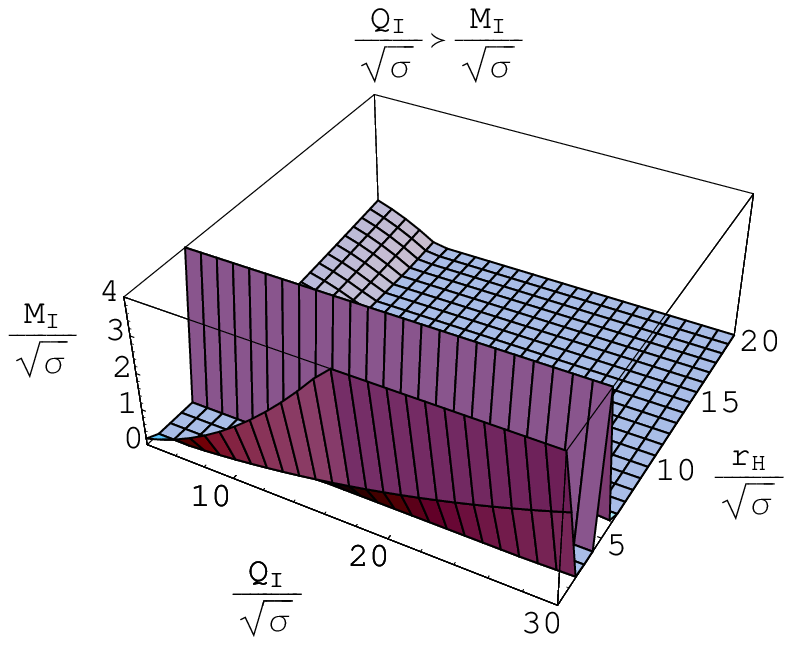, width=0.34\linewidth}\\
\end{tabular}
\caption{$\frac{M_I}{\sqrt{\sigma}}$ versus
$\frac{r_H}{\sqrt{\sigma}}$ for $t=4.00\sqrt{\sigma}$.}
\end{figure}

\begin{figure}
\begin{tabular}{c}
\epsfig{file=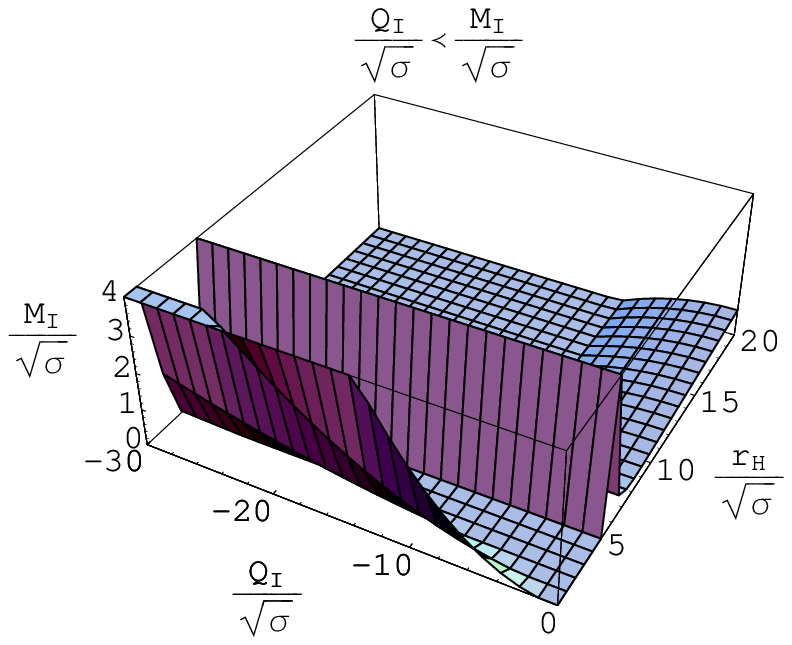, width=0.34\linewidth}
\epsfig{file=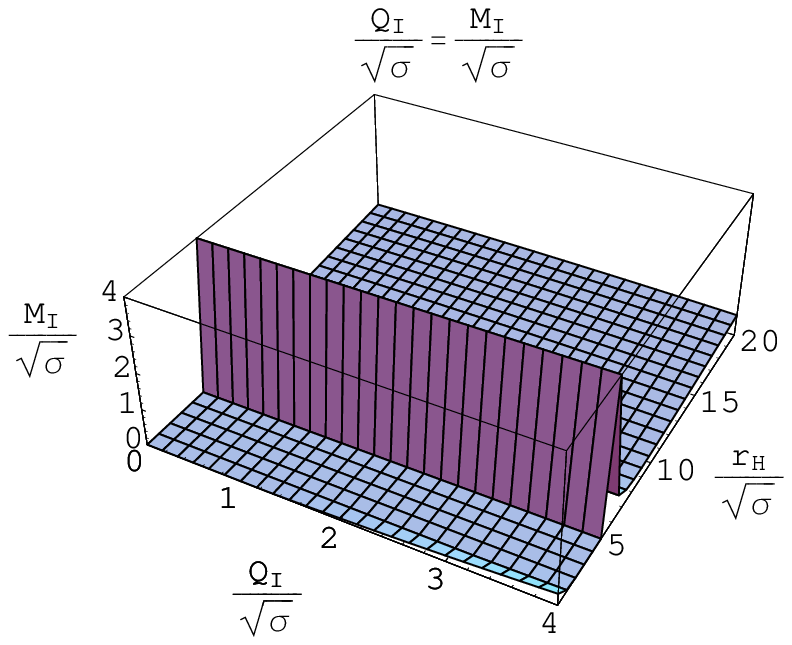, width=0.34\linewidth}
\epsfig{file=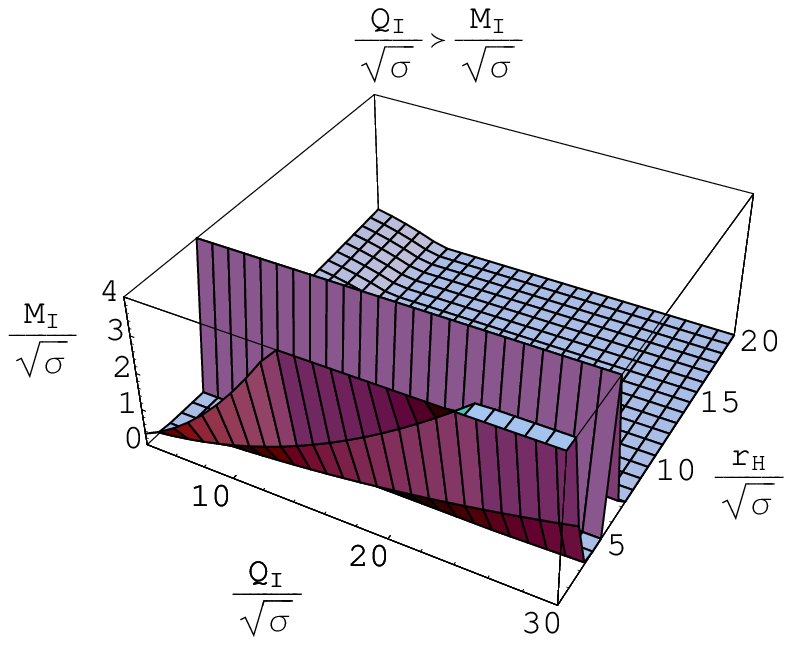, width=0.34\linewidth}
\end{tabular}
\caption{$\frac{M_I}{\sqrt{\sigma}}$ versus
$\frac{r_H}{\sqrt{\sigma}}$ for $t=5.00\sqrt{\sigma}$.}
\begin{tabular}{c}
\epsfig{file=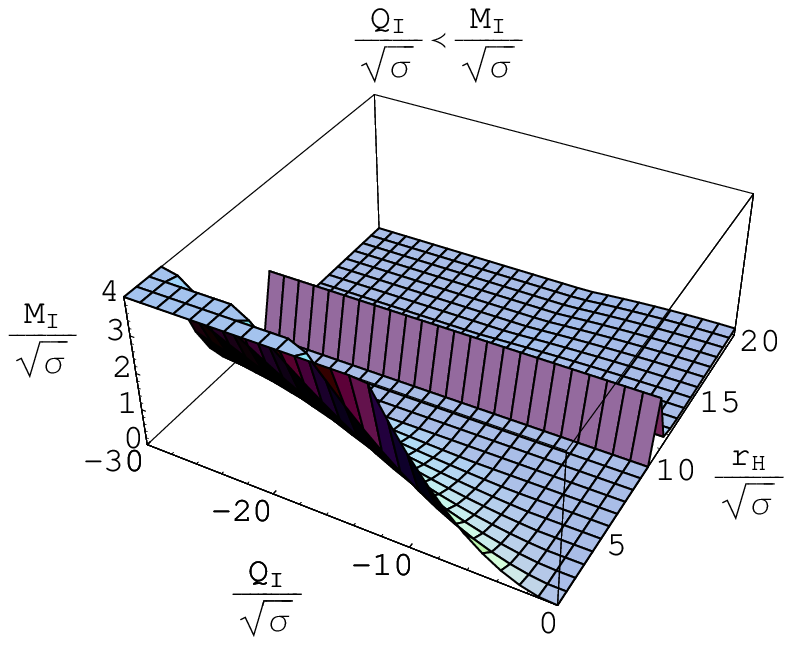, width=0.34\linewidth}
\epsfig{file=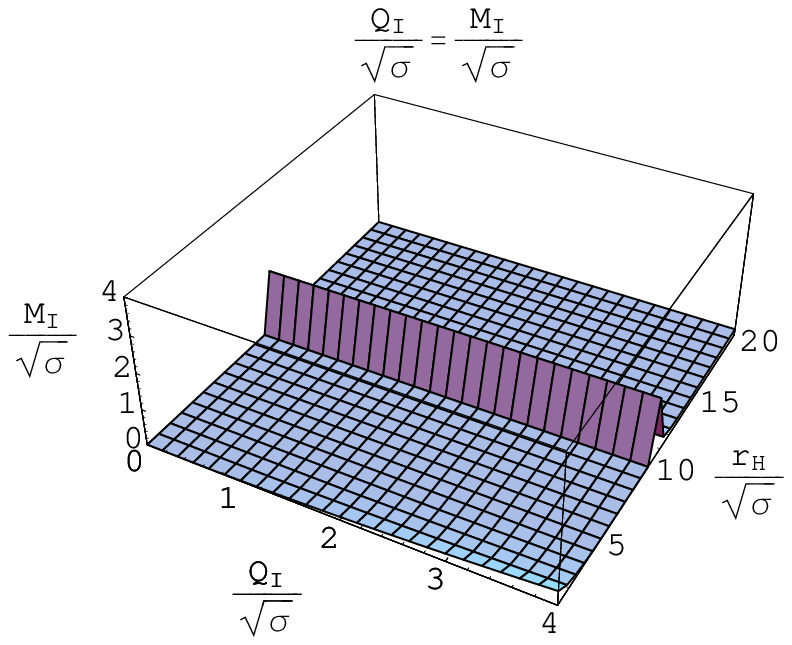, width=0.34\linewidth}
\epsfig{file=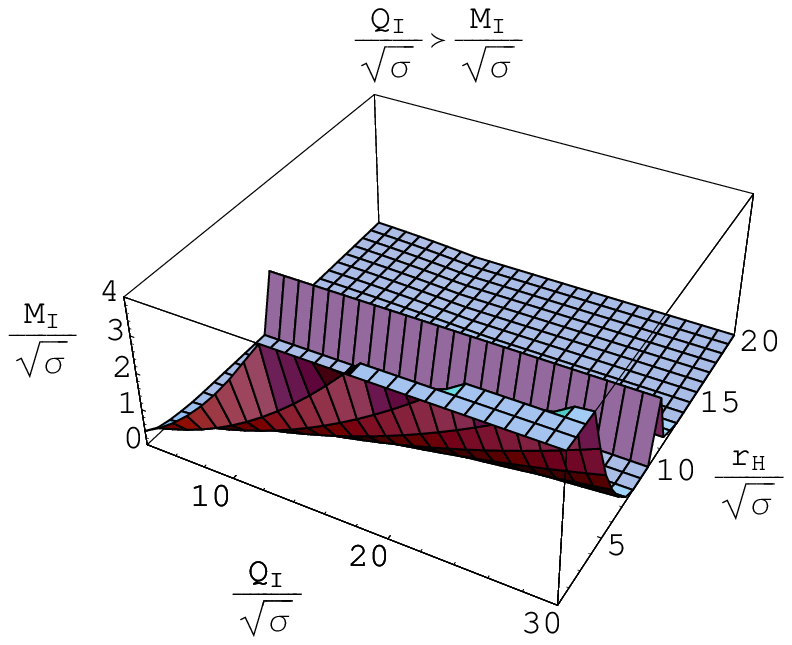, width=0.34\linewidth}\\
\end{tabular}
\caption{$\frac{M_I}{\sqrt{\sigma}}$ versus
$\frac{r_H}{\sqrt{\sigma}}$ for $t=10.00\sqrt{\sigma}$.}
\begin{tabular}{c}
\epsfig{file=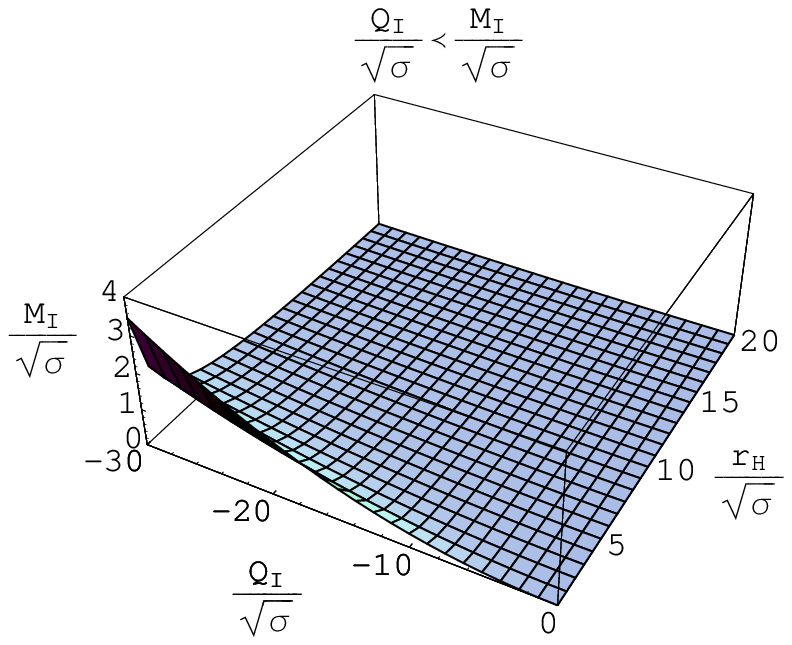, width=0.34\linewidth}
\epsfig{file=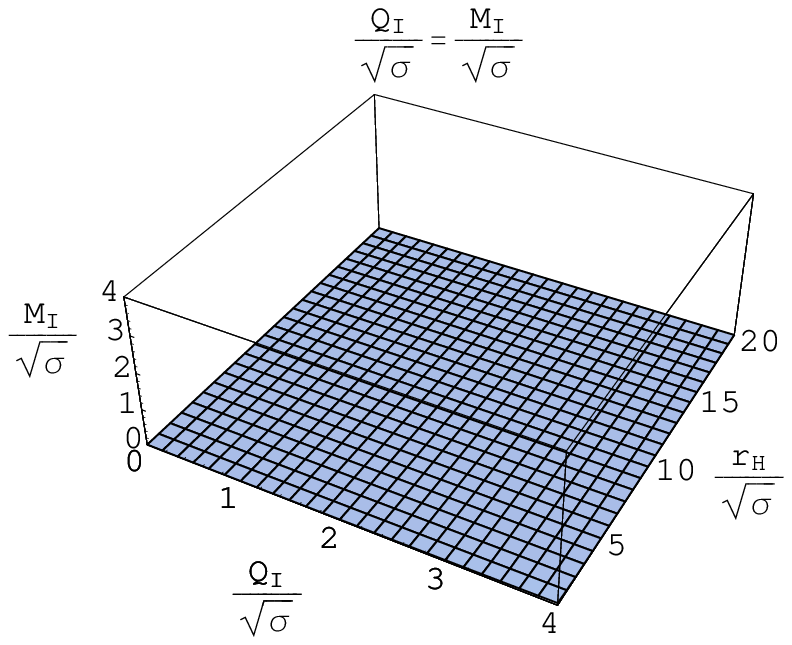, width=0.34\linewidth}
\epsfig{file=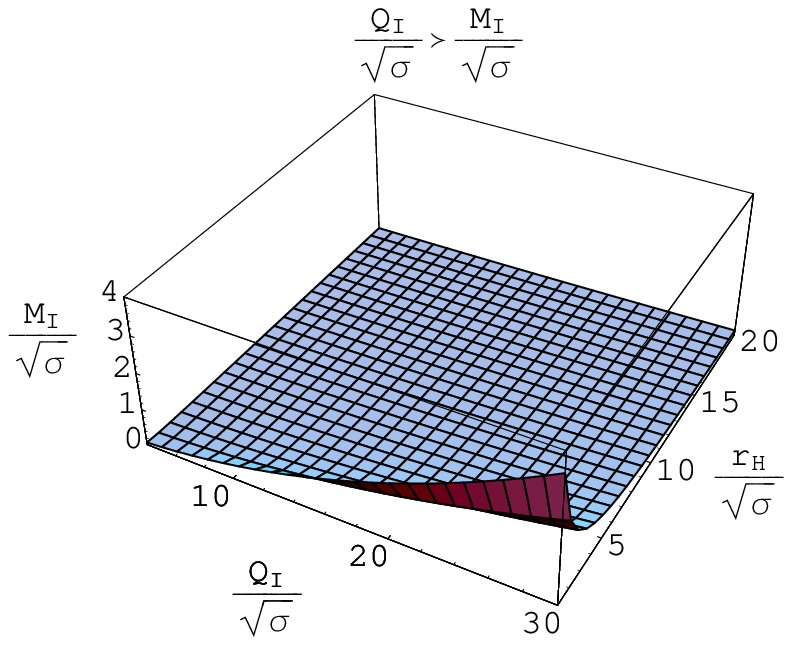, width=0.34\linewidth}
\end{tabular}
\caption{ $\frac{M_I}{\sqrt{\sigma}}$ versus
$\frac{r_H}{\sqrt{\sigma}}$ for $t=100.00\sqrt{\sigma}$.}
\end{figure}

The NC form (\ref{12}) has a coordinate singularity at the event
horizon $r_H$, i.e.,
\begin{equation}
r_H=M_\sigma(t,r_H)+\sqrt{M_\sigma^2(t,r_H)-Q_\sigma^2(t,r_H)}.\label{17}
\end{equation}
The analytical solution of this equation is not possible, however,
we can analyze the results graphically. For this purpose, we
substitute the values of $M_\sigma$ and $Q_\sigma$ from
Eq.(\ref{13.5}) in (\ref{17}) and obtain
\begin{eqnarray}
r_H^2&-&2r_HM_I\left[\varepsilon
\left(\frac{r_H-t}{2\sqrt{\sigma}}\right)
\left(1+\frac{t^2}{2\sigma}\right)-\frac{r_H}
{\sqrt{\sigma\pi}}e^{-\frac{(r_H-t)^2}{4\sigma}}\left(1+\frac{t}{r_H}\right)\right]
\nonumber\\
&=&-Q_I^2\left[\varepsilon^2\left(\frac{r_H-t}{2\sqrt{\sigma}}\right)
-\frac{r_H-t}{\sqrt{2\sigma\pi}}\varepsilon
\left(\frac{r_H-t}{\sqrt{2\sigma}}\right)\right].\label{18}
\end{eqnarray}
The graphical representation of this equation, shown in Figures
\textbf{10-17}, is consistent with Table \textbf{1}. The initial
mass (greater than the remnant mass) yields three possible causal
structures depending on different values of initial charge and
horizon radius with the passage of time.

We can summarize the behavior as follows:
\begin{itemize}
\item For $\frac{Q_I}{\sqrt{\sigma}}<\frac{M_I}{\sqrt{\sigma}}$,
$\frac{M_I}{\sqrt{\sigma}}\rightarrow 0$ as
$\frac{Q_I}{\sqrt{\sigma}}\rightarrow\infty$
and $\frac{r_H}{\sqrt{\sigma}}\rightarrow\infty$.

In this case, figures show the stable phase of the BH. As times
pases, BH starts evaporation (due to charge), its mass reduces and
approaches to zero for all horizon radius.

\item For $\frac{Q_I}{\sqrt{\sigma}}=\frac{M_I}{\sqrt{\sigma}}$,
$\frac{M_I}{\sqrt{\sigma}}\rightarrow 0$ for all
$\frac{r_H}{\sqrt{\sigma}}$ and
$\frac{Q_I}{\sqrt{\sigma}}$ at large times.

Here we see from figures the initial stage of the BH evaporation.
Black hole mass exhibits constant behavior (with the passage of
time) for small range of horizon radius indicating no effect of
charge. For large and due to effect of charge, BH mass approaches to
zero for all horizon radius.

\item For $\frac{Q_I}{\sqrt{\sigma}}>\frac{M_I}{\sqrt{\sigma}}$,
$\frac{M_I}{\sqrt{\sigma}}\rightarrow 0$ as
$\frac{Q_I}{\sqrt{\sigma}}\rightarrow 0$ and
$\frac{r_H}{\sqrt{\sigma}}\rightarrow \infty$.

This case yields the final stage of the BH evaporation. As time
progresses, BH evaporates completely, i.e., its mass and hence
temperature approaches to zero for all horizon radius.
\end{itemize}
These results imply the BH evaporation which leads to information
about the instability of the BH due to charge and hence it must
include a naked singularity. The total evaporation of the BH is
possible when we consider the time varying mass of the BH
\cite{7,21}.

\section{Hawking Radiation as Tunneling}

In this section, we examine the radiation spectrum of RN-like NC BH
by quantum tunneling \cite{8}. The tunneling is a process where a
charged particle moves in dynamical geometry and passes through the
horizon without any singularity. It provides the emission rate of
tunneled particle and depends on the key idea of energy
conservation. The mass of the BH decreases appropriately when the
virtual particle is emitted. This leads to a nonzero tunneling
amplitude, which satisfies original Hawking calculation \cite{A2}.
In this process, the coordinate system used to eliminate coordinate
singularity at the horizon, is known as Painlev$\acute{e}$
coordinate system \cite{23}. The Painlev$\acute{e}$'s time
coordinate transformation is defined as
\begin{equation}
dt\rightarrow dt-\frac{\sqrt{1-F(t,r)}}{F(t,r)}dr.\label{19}
\end{equation}
Using this transformation, the corresponding spacetime (\ref{2}) can be
written as
\begin{eqnarray}
ds^2&=&-F(t,r)dt^2+2\sqrt{1-F(t,r)}dtdr+dr^2+r^2d\Omega^2,\nonumber\\
ds^2&=&-\left(1-\frac{2M_\sigma(t,r)}{r}+\frac{Q_\sigma^2(t,r)}{r^2}\right)
dt^2\nonumber\\&+&2\sqrt{\frac{2M_\sigma(t,r)}{r}
-\frac{Q_\sigma^2(t,r)}{r^2}}dtdr+dr^2+r^2d\Omega^2.\label{20}
\end{eqnarray}
The outgoing motion (radial null geodesics, $ds^2=d\Omega^2=0$) of
the massless particles takes the form
\begin{equation}
\frac{dr}{dt}=1-\sqrt{1-F(t,r)}.\label{21}
\end{equation}
For an approximate value of $F(t,r)$ (short distances at the
neighborhood of the BH horizon), we expand $F(t,r)$ up to first
order by using Taylor series, i.e.,
\begin{equation}
F(t,r)|_t=F(t,r_H)|_t+F^\prime(t,r_H)|_t(r-r_H)+O((r-r_H)^2)|_t.\label{22}
\end{equation}
Consequently, Eq.(\ref{21}) becomes
\begin{equation}
\frac{dr}{dt}\simeq\frac{1}{2}F^\prime(t,r_H)(r-r_H)\simeq\kappa(M_I,Q_I)(r-r_H),\label{23}
\end{equation}
where $\kappa(M_I,Q_I)\simeq\frac{1}{2}F^\prime(t,r_H)$ is the
surface gravity.

Now we calculate Hawking temperature of the RN-like BH. There are
semiclassical methods to derive the Hawking temperature in the
Vaidya BH \cite{24}. Using
$T_H=\frac{\kappa}{2\pi}=\frac{1}{4\pi}F^\prime(t,r_H)|_t$, it
follows that
\begin{eqnarray}
T_H&=&\frac{1}{4\pi}\left[-2M\left\{\left(1+\frac{t^2}{2\sigma}\right)
\left(\frac{e^{-\frac{(r_H-t)^2}
{4\sigma}}}{r_H\sqrt{\pi\sigma}}-\frac{\varepsilon
\left(\frac{r_H-t}{2\sqrt{\sigma}}\right)}{r_H^2}\right)
-\frac{e^{-\frac{(r_H-t)^2}{4\sigma}}}{\sqrt{\pi\sigma}}\left(\frac{-t}{r_H^2}
\right.\right.\right.\nonumber\\
&-&\left.\left.\frac{(r_H-\frac{t^2}{r_H})}{2\sigma}\right)\right\}+Q^2
\left\{\left(
\frac{2e^{-\frac{(r_H-t)^2}{4\sigma}}}{r_H^2\sqrt{\pi\sigma}}\varepsilon
\left(\frac{r_H-t}{2\sqrt{\sigma}}\right)
-\frac{2}{r_H^3}\varepsilon^2\left(\frac{r_H-t}{2\sqrt{\sigma}}\right)
\right)\right.\nonumber\\
&-&\left.\left.\frac{1}{\sqrt{2\pi\sigma}}\left(
\left(\frac{1}{r_H}-\frac{t}{r_H^2}\right)\frac{\sqrt{2e^{-\frac{(r_H-t)^2}
{4\sigma}}}}{\sqrt{\pi\sigma}}+\varepsilon\left(\frac{r_H-t}{\sqrt{2\sigma}}\right)
\left(-\frac{1}{r_H^2}-\frac{2t}{r_H^3}\right)\right)\right\}\right].\nonumber\\&&
\label{24}
\end{eqnarray}
\begin{figure}\center
\epsfig{file=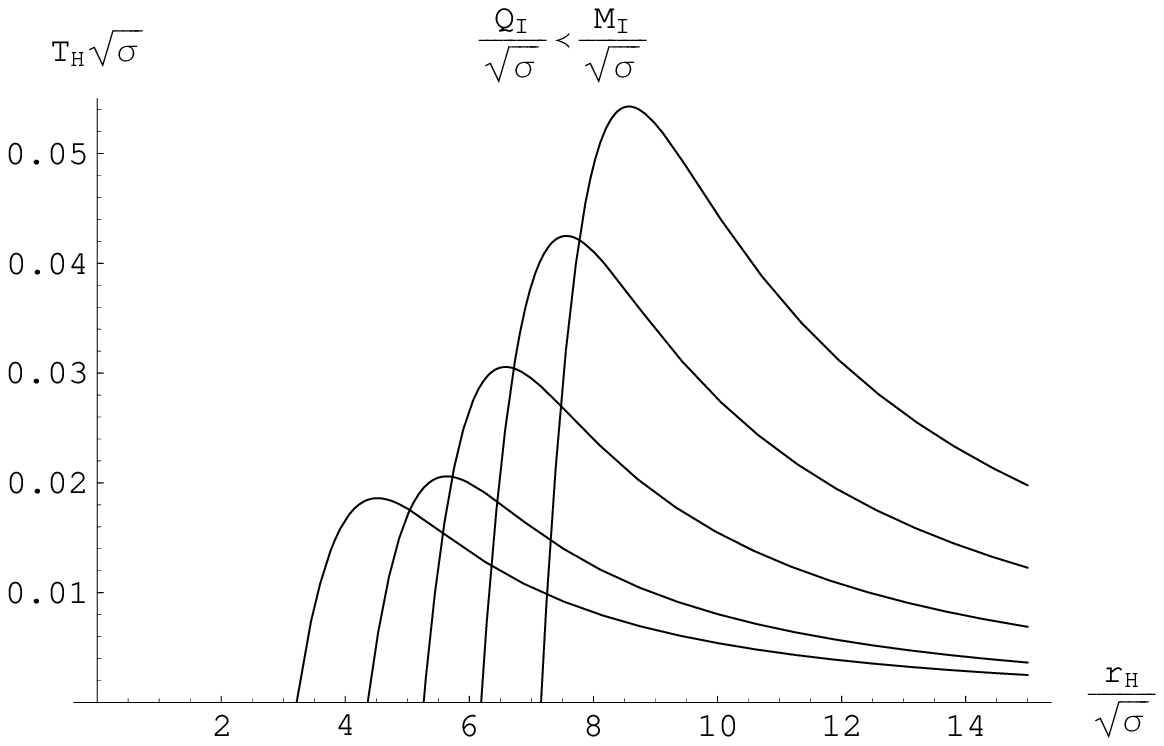, width=0.50\linewidth}\\
\caption{Here $Q_I=2.00\sqrt{\sigma} < M_I$.}
\epsfig{file=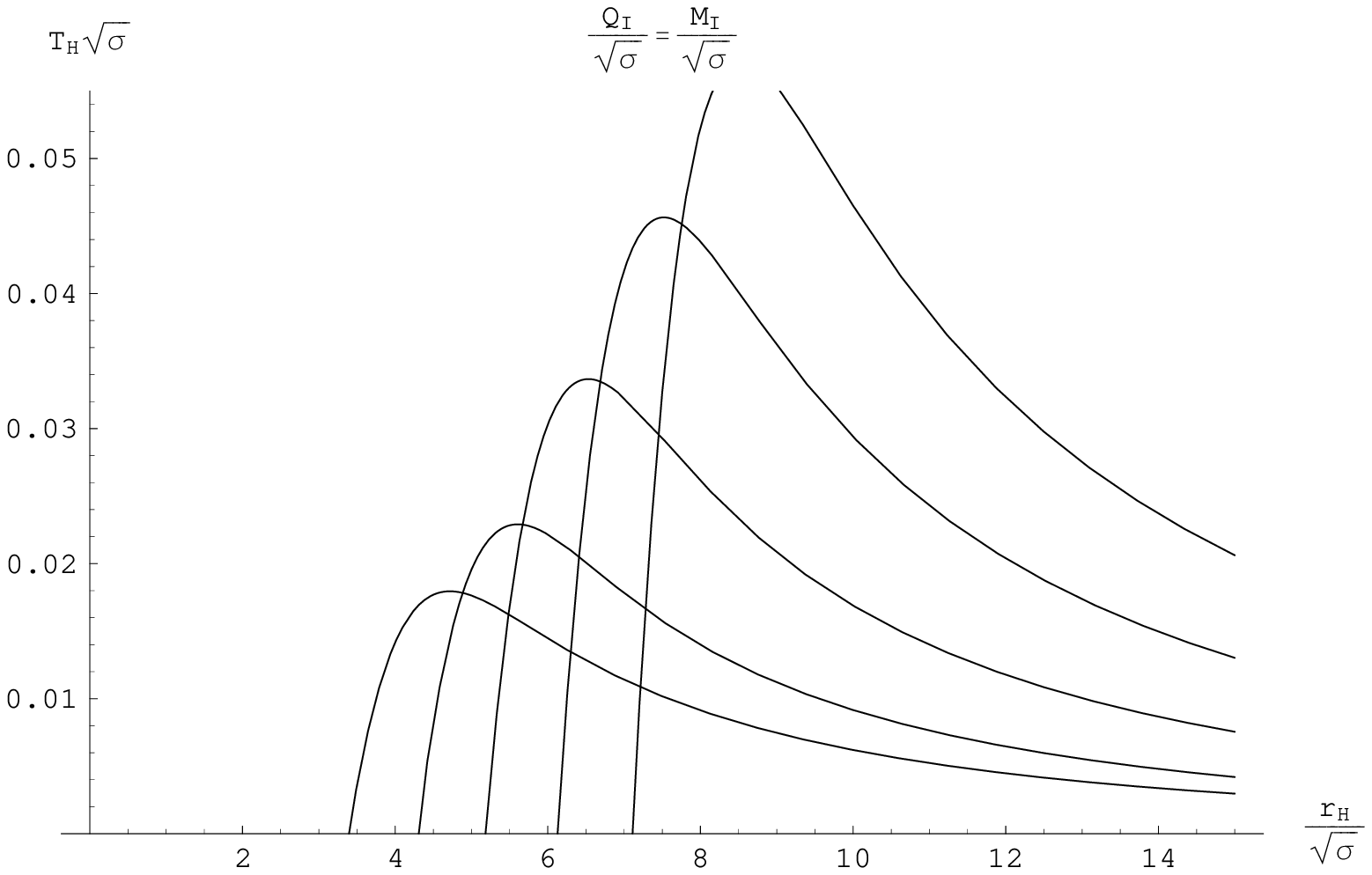, width=0.50\linewidth}\\
\caption{Here $Q_I = M_I$.} \epsfig{file=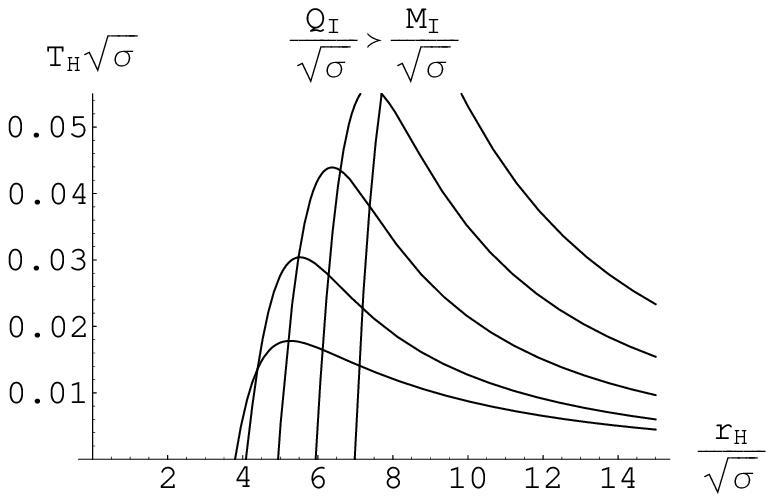,
width=0.50\linewidth} \caption{Here $Q_I=5.00\sqrt{\sigma} > M_I$.}
\end{figure}
For $t=0=Q_I$, this reduces to the Hawking temperature of
the NC Schwarzschild case \cite{3}.

Figures \textbf{18-20} show the behavior of Hawking temperature,
$T_H\sqrt{\sigma}$ versus horizon radius,
$\frac{r_H}{\sqrt{\sigma}}$ with fixed $M_I=3.00\sqrt{\sigma}$. When
BH evaporates, there is no radiation and hence temperature
approaches to zero. The graphs turn out to be smooth at the final
stage of the BH evaporation. This can also be explained as follows.
When the temperature reaches a maximum definite value at the minimal
nonzero value of the horizon radius $r_0$, then it starts to cool
down up to absolute zero and leads the mass to approach to zero. For
all the three possibilities of $\frac{M_I}{\sqrt{\sigma}}$ and
$\frac{Q_I}{\sqrt{\sigma}}$, i.e.,
$\frac{Q_I}{\sqrt{\sigma}}<\frac{M_I}{\sqrt{\sigma}}$,
$\frac{Q_I}{\sqrt{\sigma}}=\frac{M_I}{\sqrt{\sigma}}$ and
$\frac{Q_I}{\sqrt{\sigma}}>\frac{M_I}{\sqrt{\sigma}}$, the graphs of
Hawking temperature give the following behavior.
\begin{itemize}
\item For $\frac{Q_I}{\sqrt{\sigma}}<\frac{M_I}{\sqrt{\sigma}}$,
the behavior of curves are the same as for the Schwarzschild case
\cite{P2}.
\item For $\frac{Q_I}{\sqrt{\sigma}}=\frac{M_I}{\sqrt{\sigma}}$,
temperature increases at minimal horizon radius.
\item For $\frac{Q_I}{\sqrt{\sigma}}>\frac{M_I}{\sqrt{\sigma}}$,
horizon radius changes its position with increasing temperature.
\end{itemize}

Now we would like to discuss the effect of electromagnetic field on
the emission rate of charged particles tunnel through the quantum
horizon of the BH. Here we assume that an electromagnetic field is
present outside the BH. The Lagrangian function for such a Maxwell
gravity system can be defined as
\begin{equation}
L=L_{matt}+L_{el},\label{}
\end{equation}
where $L_{el}=-\frac{1}{4}F_{ab}F^{ab}$ is the Maxwell Lagrangian
function and $F^{ab}$ is the Maxwell field tensor given by
\begin{equation}
F^{ab}=\partial^a\Phi^b-\partial^b\Phi^a,\label{}
\end{equation}
where $\Phi_a=(\Phi,0,0,0)$ is the electromagnetic 4-potential. The
action and the rate of emission of a particle in tunneling process
are, respectively, defined as \cite{P3}
\begin{equation}
I=\int_{t_{in}}^{t_{out}}(L_{matt}-p_\Phi\dot{\Phi})dt,\quad
\Gamma\sim\exp({\textmd{-2Im}\textit{I}}),\label{25}
\end{equation}
where $p_\Phi$ is the canonical momentum conjugate to $\Phi$. In the
tunneling process, the imaginary part of the amplitude for an
$s$-wave, representing the outgoing positive energy particles which
cross the horizon outward from $r_{in}$ to $r_{out}$, is given by
\begin{equation}
{\textmd{Im}\textit{I}}=\textmd{Im}\int_{r_{in}}^{r_{out}}\left(p_r-
\frac{p_\Phi\dot{\Phi}}{\dot{r}}\right)dr=\textmd{Im}\int_{r_{in}}^{r_{out}}
\left[\int_{(0,0)}^{(p_r,p_\Phi)}dp^\prime_r-\frac{\dot{\Phi}}{\dot{r}}
dp^\prime_\Phi\right]dr.\label{26}
\end{equation}
Hamilton's equations of motion,
\begin{eqnarray}
\frac{dr}{dt}&=&\frac{dH}{dp_r}|_{(r;\Phi,p_\Phi)}=\frac{d(M-E)}{dp_r}=-\frac{dE}{dp_r},\nonumber\\
\frac{d\Phi}{dt}&=&\frac{dH}{dp_\Phi}|_{(\Phi;r,p_r)}=\Phi(Q-q)\frac{dq}{dp_\Phi},
\end{eqnarray}
provide the following relation of momentum and energy,
\begin{equation}
{\textmd{Im}\textit{I}}=\textmd{Im}\int_{r_{in}}^{r_{out}}
\left[\int_{(0,0)}^{(H,q)}\frac{dH^\prime}{\dot{r}}-
\frac{\Phi(Q-q^\prime)}{\dot{r}}dq^\prime\right]dr.\label{27}
\end{equation}

In this process, particles and antiparticles can be described as a
positive and negative energy solution of the wave equation
respectively. Black hole accretes a small negative energy, which
decreases its mass. Replacing $M_I$ by $M_I-E$, $Q_I$ by $Q_I-q$ and
substituting Eq.(\ref{23}) in (\ref{27}), we get
\begin{eqnarray}
{\textmd{Im}\textit{I}}&=&-\textmd{Im}\int_{r_{in}}^{r_{out}}
\left[\int_{(0,0)}^{(E,q)}\frac{dE^\prime}
{\kappa(M_I-E^\prime,Q_I-q^\prime)(r-r_H)}\right.\nonumber\\
&+&\left.\frac{\Phi(Q-q^\prime)dq^\prime}
{\kappa(M_I-E^\prime,Q_I-q^\prime)(r-r_H)}\right]dr.\label{28}
\end{eqnarray}
This integral has a pole at the horizon $r_H$. To avoid this
pole, we perform contour integration with the condition
$r_{in}>r_{out}$ and obtain
\begin{eqnarray}
{\textmd{Im}\textit{I}}&=&-\textmd{Im} \left[\int_{(0,0)}^{(E,q)}
\frac{dE^\prime}{\kappa(M_I-E^\prime,Q_I-q^\prime)}+
\frac{\Phi(Q-q^\prime)dq^\prime}
{\kappa(M_I-E^\prime,Q_I-q^\prime)}\right]\nonumber\\
&\times&\int_{r_{in}}^{r_{out}}\frac{dr}{r-r_H}\nonumber\\
&=&\pi\left[\int_{(0,0)}^{(E,q)}
\frac{dE^\prime}{\kappa(M_I-E^\prime,Q_I-q^\prime)}+
\frac{\Phi(Q-q^\prime)dq^\prime}
{\kappa(M_I-E^\prime,Q_I-q^\prime)}\right].\label{29}
\end{eqnarray}
This shows that the particles rate of emission is proportional to
the surface gravity.

Using the first law of BH thermodynamics, $dM=TdS-\Phi dQ$, the
imaginary part of the action ${\textmd{Im}\textit{I}}$ is given by
\cite{30}
\begin{equation}
{\textmd{Im}\textit{I}}=-\frac{1}{2}\int_{S_{NC}(M,Q)}^{S_{NC}(M-E,Q-q)}dS
=-\frac{1}{2}\Delta S_{NC},\label{30}
\end{equation}
where $S_{NC}$ is the entropy of the NC BH while $\Delta{S_{NC}}$ is
the difference in BH entropies before and after emission. At high
energies, the tunneling amplitude (emission rate) depends on the
final and initial number of microstates available to the system
\cite{31}-\cite{34} implying that the emission rate is proportional
to $\exp(\Delta S_{NC})$, i.e.,
\begin{equation}
\Gamma\sim\frac{e^{S_{final}}}{e^{S_{initial}}}=e^{\Delta S_{NC}}=
e^{S_{NC}(M_I-E,Q_I-q)-S_{NC}(M_I,Q_I)}.\label{31}
\end{equation}
It follows that the emission spectrum cannot be precisely thermal.
The modified NC tunneling amplitude $\Gamma$ can be computed if we
know the analytic form of $\exp(\Delta {S_{NC}})$.

According to quantum theory, a BH is neither an absolute stationary
state nor even a relative stationary state, it is an excited state
of gravity. Vacuum state (excited state) generates the spontaneous
emission of virtual particles. Thus the emission of charged
particles by BH is physically equivalent to the spontaneous emission
by an excited state \cite{A3}.

\section{Summary}

In this paper, first of all, we have derived spherically symmetric
charged Vaidya metric in RN-like form and its NC version. Noncommutativity
implies a minimal nonzero mass that allows the existence of an event horizon. In order
to investigate the BH horizon radius depending on time, mass and
charge, we have examined the behavior of $F(t,r)$ in the form of
graphs, shown in Figures \textbf{1-9} for three possible
structures:$~$(i)$~M_I>M_0,~$(ii)$~M_I=M_0,~$(iii)$~M_I<M_0$. These
have further been discussed for three possibilities of initial mass
and initial charge, i.e., $\frac{Q_I}{\sqrt{\sigma}}
<\frac{M_I}{\sqrt{\sigma}}$,
$\frac{Q_I}{\sqrt{\sigma}}=\frac{M_I}{\sqrt{\sigma}}$ and
$\frac{Q_I}{\sqrt{\sigma}}>\frac{M_I}{\sqrt{\sigma}}$. The first
case provides two different possible horizons. Case (ii) represents
the possibility of an extremum structure with one degenerate event
horizon with time in the presence of charge. The last case shows
that curves do not indicate any event horizon.

In Figures \textbf{10-17}, the effects of charge on the BH
evaporation are shown. The relationships between mass and charge indicate
three different stages of BH mass and charge which lead to evaporation of the BH.
Using Table \textbf{1}, we have found that BH
mass approaches to zero as horizon radius tend to
infinity with time. This shows that structure of stable BH remnant
having capability to store information has been failed and
information would disappear from our world. Hence, this leads to the
evaporation of the BH and the final phase is a naked singularity. We have found that
BH evaporates completely in the large time limit. We also see from
these figures that the cases
$\frac{Q_I}{\sqrt{\sigma}}<\frac{M_I}{\sqrt{\sigma}}$ and
$\frac{Q_I}{\sqrt{\sigma}}>\frac{M_I}{\sqrt{\sigma}}$ indicate
reverse behavior of each other.

The analysis of Hawking temperature (Figures \textbf{18-20}) shows
similar behavior as that of the Schwarzschild spacetime. In the
presence of charge, temperature attains a maximum position at the
minimal nonzero horizon radius. As horizon radius increases,
temperature vanishes which corresponds to the BH evaporation, i.e.,
mass approaches to zero. Finally, we have discussed the Hawking
radiation by using Parikh-Wilczek tunneling process through the
quantum horizon. The emission rate has been found consistent with
the unitary theory. We have extended this analysis to compute the
tunneling amplitude of charged massive particles from the RN-like
Vaidya BH. It is mentioned here that corrections due to NC can be
considered before the BH mass approaches to the Planck mass.

It would be interesting to extend this work to the dyadosphere of a
RN solution and the regular BH solutions in NC space. It would also
be worthwhile to examine the behavior of thermodynamical quantities,
evaporation of BH remnant and Hawking radiation as tunneling for
these solutions.

\newpage

{\bf Acknowledgment}

\vspace{0.25cm}

We would like to thank the Higher Education Commission, Islamabad,
Pakistan for its financial support through the {\it Indigenous
Ph.D. 5000 Fellowship Program Batch-IV}.

\end{document}